\documentclass[%
pra, twocolumn,
 amsmath,amssymb,
 aps,
]{revtex4-2}

\usepackage{graphicx}
\usepackage{dcolumn}
\usepackage{bm}

\usepackage{amsmath, amsthm, amssymb,ulem}
\usepackage{float,epsfig}
\usepackage{color}
\usepackage{multirow}
\usepackage{cleveref}
\usepackage{braket}
\usepackage{subfigure}

\begin{document}

\newtheorem{theorem}{\bf Theorem}[section]
\newtheorem{proposition}[theorem]{\bf Proposition}
\newtheorem{definition}[theorem]{\bf Definition}
\newtheorem{corollary}[theorem]{\bf Corollary}
\newtheorem{example}[theorem]{\bf Example}
\newtheorem{exam}[theorem]{\bf Example}
\newtheorem{remark}[theorem]{\bf Remark}
\newtheorem{lemma}[theorem]{\bf Lemma}
\newtheorem{statement}[theorem]{\bf Statement}
\newcommand{\nrm}[1]{|\!|\!| {#1} |\!|\!|}

\newcommand{\calL}{{\mathcal L}}
\newcommand{\calX}{{\mathcal X}}
\newcommand{\calA}{{\mathcal A}}
\newcommand{\calB}{{\mathcal B}}
\newcommand{\calC}{{\mathcal C}}
\newcommand{\calK}{{\mathcal K}}
\newcommand{\C}{{\mathbb C}}
\newcommand{\R}{{\mathbb R}}
\newcommand{\U}{{\mathrm U}}
\renewcommand{\SS}{{\mathbb S}}
\newcommand{\LL}{{\mathbb L}}
\newcommand{\st}{{\star}}
\def\kernel{\mathop{\rm kernel}\nolimits}
\def\sigan{\mathop{\rm span}\nolimits}

\newcommand{\klasse}{{\boldsymbol \Delta}}

\newcommand{\ba}{\begin{array}}
\newcommand{\ea}{\end{array}}
\newcommand{\von}{\vskip 1ex}
\newcommand{\vone}{\vskip 2ex}
\newcommand{\vtwo}{\vskip 4ex}
\newcommand{\dm}[1]{ {\displaystyle{#1} } }

\newcommand{\be}{\begin{equation}}
\newcommand{\ee}{\end{equation}}
\newcommand{\beano}{\begin{eqnarray*}}
\newcommand{\eeano}{\end{eqnarray*}}
\newcommand{\inp}[2]{\langle {#1} ,\,{#2} \rangle}
\def\bmatrix#1{\left[ \begin{matrix} #1 \end{matrix} \right]}
\def \noin{\noindent}
\newcommand{\evenindex}{\Pi_e}

\newcommand{\tb}[1]{\textcolor{blue}{ #1}}
\newcommand{\tm}[1]{\textcolor{magenta}{ #1}}
\newcommand{\tre}[1]{\textcolor{red}{ #1}}



\def \K{{\mathbf k}}
\def \N{{\mathbb N}}
\def \R{{\mathbb R}}
\def \F{{\mathbb F}}
\def \C{{\mathbb C}}
\def \Q{{\mathbb Q}}
\def \Z{{\mathbb Z}}
\def \I{{\mathbb I}}
\def \D{{\mathcal D}}
\def \H{{\mathcal H}}
\def \P{{\mathcal P}}
\def \M{{\mathcal M}}
\def \B{{\mathcal B}}
\def \O{{\mathcal O}}
\def \calG{{\mathcal G}}
\def \PO{{\mathcal {PO}}}
\def \X{{\mathcal X}}
\def \Y{{\mathcal Y}}
\def \calW{{\mathcal W}}
\def \pf{{\bf Proof: }}
\def \lam{{\lambda}}
\def\lc{\left\lceil}   
\def\rc{\right\rceil}
\def \N{{\mathbb N}}
\def \Ls{{\Lambda}_{m-1}}
\def \Gb{\mathrm{G}}
\def \Hb{\mathrm{H}}
\def \Delta{\triangle}
\def \Rar{\Rightarrow}
\def \p{{\mathsf{p}(\lam; v)}}

\def \D{{\mathbb D}}

\def \tr{\mathrm{Tr}}
\def \cond{\mathrm{cond}}
\def \lam{\lambda}
\def \sig{\sigma}
\def \sign{\mathrm{sign}}

\def \ep{\epsilon}
\def \diag{\mathrm{diag}}
\def \rev{\mathrm{rev}}
\def \vec{\mathrm{vec}}

\def \ham{\mathsf{Ham}}
\def \herm{\mathsf{Herm}}
\def \sym{\mathsf{sym}}
\def \odd{\mathsf{sym}}
\def \en{\mathrm{even}}
\def \rank{\mathrm{rank}}
\def \pf{{\bf Proof: }}
\def \dist{\mathrm{dist}}
\def \rar{\rightarrow}

\def \rank{\mathrm{rank}}
\def \pf{{\bf Proof: }}
\def \dist{\mathrm{dist}}
\def \Re{\mathsf{Re}}
\def \Im{\mathsf{Im}}
\def \re{\mathsf{re}}
\def \im{\mathsf{im}}

\def \sym{\mathsf{sym}}
\def \sksym{\mathsf{skew\mbox{-}sym}}
\def \odd{\mathrm{odd}}
\def \even{\mathrm{even}}
\def \herm{\mathsf{Herm}}
\def \skherm{\mathsf{skew\mbox{-}Herm}}
\def \str{\mathrm{ Struct}}
\def \eproof{$\blacksquare$}

\def \bS{{\bf S}}
\def \cA{{\cal A}}
\def \E{{\mathcal E}}
\def \X{{\mathcal X}}
\def \F{{\mathcal F}}
\def \cH{\mathcal{H}}
\def \cJ{\mathcal{J}}
\def \tr{\mathrm{Tr}}
\def \range{\mathrm{Range}}
\def \adj{\star}

\preprint{APS/123-QED}


\title{Stronger resilience to disorder in 2D quantum walks than in 1D}

\author{Amrita Mandal}
\email{mandalamrita55@gmail.com}
\author{Ujjwal Sen}
\email{ujjwalsen0601@gmail.com}

\affiliation{Harish-Chandra Research Institute, A CI of Homi Bhabha National Institute, Chhatnag Road, Jhunsi,  Prayagraj 211 019, India}


\begin{abstract}

We study the response of spreading behavior, of two-dimensional discrete-time quantum walks, to glassy disorder in the jump length. We consider different discrete probability distributions to mimic the disorder, and three types of coin operators, viz., Grover, Fourier, and Hadamard, to analyze the scale exponent of the disorder-averaged spreading. We find that the ballistic spreading of the clean walk is inhibited in presence of disorder, and the walk becomes sub-ballistic but remains super-diffusive. The resilience to disorder-induced inhibition is stronger in two-dimensional walks, for all the considered coin operations, in comparison to the same in one dimension. The quantum advantage of quantum walks is therefore more secure in two dimensions than in one.

\end{abstract}

\keywords{ Quantum walk, Grover matrix}
\maketitle



\section{Introduction}\label{sec:intro}
For the past few decades, the area of quantum walks is among the hotly pursued research topics in quantum services, and is known to provide a universal quantum computation framework \cite{childs2009universal,childs2013universal}.  The intriguing behavior of quantum ``particles'' makes quantum walks significantly different from that of its classical analog, the classical random walks (CRWs). The altered spreading of quantum walks provides a useful tool for developing quantum algorithms with fundamentally faster computation than in the classical regime \cite{ambainis2003quantum,childs2003exponential,shenvi2003quantum,ambainis2007quantum}.  
  Quantum walks have been argued to play crucial roles in physical processes such as electric-field driven systems, photosynthesis, protein DNA target search, and more \cite{oka2005breakdown,mohseni2008environment,d2021protein}. They are useful to simulate a wide range of quantum phenomena such as relativistic quantum dynamics \cite{di2013quantum}, topological phases \cite{kitagawa2010exploring,obuse2011topological,mittal2021persistence}, and neutrino oscillations \cite{mallick2017neutrino}.
 Quantum walks are defined for both continuous and discrete time-steps \cite{meyer1996kt, farhi1998quantum, venegas2012quantum}. In this article, we focus on discrete-time (coined) quantum walks (DTQWs) \cite{aharonov2001quantum,ambainis2001one,kempe2003quantum,konno2002quantum,venegas2008quantum}.

 A DTQW on a graph or lattice propagates with the application of a coin operator followed by a conditional shift operator to the state vector, at each time step. In general, for DTQW (without disorder), the shift operator or displacement operator conditioned on the coin state displaces the particle one unit through the lattice, from its current position, to an adjacent vertex, and all the adjacent vertices are typically covered via conditioning on different coin states. The coin and shift operators perform 
 uniformly with the same set of outputs, relative to the input of that step, at all time steps. 
 
 
Disorder is ubiquitous in physical systems, 
and quantum walks with disorder have been studied for disorder in the displacement operations \cite{lavivcka2011quantum,pires2019multiple,mukhopadhyay2020persistent,pires2020quantum,naves2022enhancing}, as well as in coin operations \cite{salimi2012asymptotic,vieira2013dynamically,rohde2013quantum,vieira2014entangling,di2016discrete,montero2016classical,wang2018dynamic,orthey2019weak,singh2019accelerated,buarque2019aperiodic,pires2021negative,naves2023quantum}. 
For example, inhomogeneity or disorder in the position-dependent coin operator yields an unusual change in the spreading behavior, so that the walk remains bounded in a certain region for all time \cite{linden2009inhomogeneous}. 
In real-life scenarios or in real experiments, it is expected that disorder may appear in quantum walks due to the unavoidable interaction between the system and the environment, or due to accidental defects in the system, or even due to engineered ones. Effects of disorder and decoherence have been studied more extensively in one-dimensional lattices than in two-dimensional ones \cite{brun2003quantum,oliveira2006decoherence,gonulol2009decoherence,chandrashekar2013decoherence,annabestani2010decoherence,PhysRevA.98.032104,yang2021decoherence,tude2022decoherence}, and even in 2D lattices the analysis is usually restricted to phase-defect disorder in shift operations \cite{xue2015localized,di2018elephant,naves2022quantum}.
In this work, we investigate the effects of disorder in the displacement operation caused by random jumps, with a fixed distribution, of the quantum walker moving from one vertex to another in a two-dimensional regular lattice.

There are broadly two types of disorders  employed 
in classical or quantum random walks: static and dynamic \cite{chatterjee1994effective,yin2008quantum,nosrati2021readout}. The modification of lattice due to the presence of impurity often shows up as static disorder for the ``particles'' in the lattice, and the time-dependent vibrations of lattice particles may lead to dynamic disorder.
Static disorder in random walks is modeled by using a time-independent probability of moving from one vertex to a neighboring one, whereas in dynamic disorder, the same is time-dependent, possibly non-deterministically. 
The transition probability per unit time in the master equation of a CRW defined on a medium with dynamic disorder is a time-dependent random process. The presence of disorders in a quantum walk randomizes the latter by providing a decoherence channel to the system and turns the walk towards the classical one by hindering the walker's spreading. For example, in 1D-DTQW realized using non-perfect optical multi-ports, dynamic disorder leads to a Gaussian-like distribution \cite{lavivcka2011quantum}. This article mainly focuses on the effects of non-deterministic dynamic disorder on 2D-DTQWs.




CRWs in 1D and 2D lattices receive special attention due - at least in part - to the celebrated result that the walker returns to the origin infinitely often with unit probability, and is referred to as the recurrence property of the walk \cite{polya1921aufgabe,revesz2013random}. The spreading, quantified by the standard deviation,  proceeds as the square root of the number of time steps in both one- and two-dimensional CRWs. In DTQWs without disorder on a line or on a square grid, there is a quadratic speedup in the spreading rate, meaning that the standard deviation progresses linearly with time steps, which predicts a faster-than-classical search algorithm \cite{nayak2000quantum,ambainis2001one,portugal2013quantum}. Thus time-scaled behavior of quantum walk distributions provides valuable insights, enabling the utilization of quantum walks for various applications in quantum technologies.

In this paper, we consider a disorder-induced shift operator and study the behavior of four-state DTQWs on the two-dimensional square lattice with four-dimensional Grover, Fourier, and Hadamard coins. Disorder in the shift operator of such walks are assumed here to come from the presence of ``glassy'' disorder, which has also been referred to in the literature as ``quenched'' disorder. We analyze the spreading of the walker by deriving the disorder-averaged scaling of the position-probability distribution. We consider several discrete probability distributions for the disorder, including Poisson, as well as certain sub- and super-Poisson distributions. For all such cases, we find that, similar to 1D-DTQWs,  spread of the quantum walker through the 2D lattice is inhibited in the presence of disorder. This signifies a transition from the ballistic quantum regime towards the diffusive classical one due to the onset of disorder. We show that there is an inverse relation between the disorder strength and the spreading rate. However, the rate - with respect to disorder strength - of loss of ballistic quantumness in the spreading dynamics of two-dimensional Grover, Fourier, and Hadamard walks is slower than that in the one-dimensional case, for the same disorder strength. 

The organization of the paper is as follows. Section~\ref{sec:DTQW} formally - though, briefly - describes 2D-DTQWs, and the different coin operators that we undertake for disorder infliction analysis. In Section~\ref{sec:DTQW disorder}, we introduce 2D-DTQWs with glassy disorder in the shift operation. In this context, we recapitulate different kinds of discrete-probability distributions, including Poisson, binomial, hypergeometric, etc., which are utilized to induce disorder in the jump lengths of the walks along the lattice. In Section~\ref{sec:analysis}, we focus on dynamic disorder and provide  systematic computational study on the spread - and its scaling - of such disordered DTQWs. Section~\ref{sec:static} briefly highlights the effects of static disorder on the same lattice. We conclude the article in Section~\ref{sec:conclusion}.


\section{Discrete-time quantum walks on two-dimensional lattice} \label{sec:DTQW}
In this section, we review DTQWs on the two-dimensional square lattice \cite{inui2004localization,vstefavnak2008recurrence}. DTQWs are defined on the Hilbert space $\H=\H_p\otimes\H_c,$
 where $\H_p$ and $\H_c$ denote the position and coin spaces, respectively.  Let \begin{equation*}
    Z_t=\{ (x, y)\in \mathbb{Z}^2 : -t\leq x\leq t, -t\leq y\leq t\}
\end{equation*} denote the square lattice with $(2t+1)^2$ vertices. The position state vector $\ket{x,y}\in \H_p$ for the position $(x,y) \in Z_t$, and after $t$ time steps, $\H_p=\mathrm{Span}\{\ket{x,y}|(x,y)\in Z_t\}$ is of dimension $(2t+1)^2.$  
Unlike the classical case, the quantum coin state can be in superpositions of the four (basis) states correspond to right, left, up, and down directions. Hence, the dimension of $\H_c$ is four. The vectors of the standard basis assigned to each of these four displacement instructions span the space $\H_c$ i.e. $\H_c=\mathrm{Span}\{\ket{0},\ket{1},\ket{2},\ket{3}\},$ where $\{\ket{l}|~l=0,1,2,3\}$ is the canonical ordered basis of $\mathbb{C}^4$. 
  Thus the total state space $\H=\mathrm{Span}\{\ket{x,y}\otimes\ket{{l}}|~(x,y)\in Z_t,l\in\{0,1,2,3\}\}$ and is isomorphic to $ \C^{(2t+1)^2}\otimes\C^4$.

 The evolution of the proposed DTQW is governed by repeated application of the unitary operator $\U=S(I_{(2t+1)^2}\otimes C)$ to the initial state of the walker and coin, where $S$ is the shift operator, $C$ is the coin operator, and $I_n$ is the identity matrix of order $n.$ For brevity, sometimes we use only $I$ to denote the identity matrix.

The conditional shift operator $S$ is given as follows:
 \beano
 S = \sum_{(x,y)\in Z_{t-1}} &&\Big( \ket{x+1, y}  \bra{x ,y}\otimes \ket{0}\bra{0}+\\
 && \phantom{ll}\ket{x-1, y}\bra{x,y}\otimes \ket{1}\bra{1}+ \\
 && \phantom{ll}\ket{x,y+1}\bra{x,y}\otimes \ket{2}\bra{2}+\\ && \phantom{ll}\ket{x,y-1}\bra{x,y}\otimes \ket{3}\bra{3}\Big).
 \eeano
 For a walker beginning at $(0,0),$ the shift operator can be applied only $t$ times.
 The coin operator $C$ acting on $\H_c$ can be any unitary operator on $\C^4$ for two-dimensional four-state DTQWs. Here, we take the three extensively used coins, viz. Grover, Hadamard, and Fourier matrices, for DTQWs defined on the two-dimensional square lattice  \cite{oliveira2006decoherence,gonulol2009decoherence}. The Grover matrix or Grover diffusion matrix for 2D-DTQWs, takes the form, $G=\frac{1}{2}\boldsymbol{1}_4\boldsymbol{1}_4^\dagger - I_4,$ where $\boldsymbol{1}_4=\frac{1}{2}\sum_{i=0}^3\ket{i}$ denotes the all-one (column) vector of dimension $4$ and $I_4$ is the identity matrix of order $4$ \cite{grover1996fast}, and the corresponding DTQW is called the Grover walk \cite{mackay2002quantum,inui2004localization,di2011mimicking}. The action of the Grover coin on basis elements will be clear from the following:
 \beano G&: &\ket{0}\rightarrow (-\ket{0}+\ket{1}+\ket{2}+\ket{3})/2, \\
 && \ket{1}\rightarrow \phantom{-}(\ket{0}-\ket{1}+\ket{2}+\ket{3})/2,\\
  && \ket{2}\rightarrow \phantom{-}(\ket{0}+\ket{1}-\ket{2}+\ket{3})/2,\\
   && \ket{3}\rightarrow \phantom{-}(\ket{0}+\ket{1}+\ket{2}-\ket{3})/2.
 \eeano
 The Fourier matrix, $F,$ corresponding to the discrete Fourier transformation, is also known as the generalized quantum Fourier transform or generalized Hadamard gate. $F$ is described as follows:
  \beano F&: &\ket{0}\rightarrow (\ket{0}+\phantom{i}\ket{1}+\ket{2}+\phantom{i}\ket{3})/2, \\
 && \ket{1}\rightarrow (\ket{0}+i\ket{1}-\ket{2}-i\ket{3})/2,\\
  && \ket{2}\rightarrow (\ket{0}-\phantom{i}\ket{1}+\ket{2}-\phantom{i}\ket{3})/2,\\
   && \ket{3}\rightarrow (\ket{0}-i\ket{1}-\ket{2}+i\ket{3})/2.
 \eeano
 DTQW with $F$ as the coin operator is named as the Fourier walk. The (single-qubit) Hadamard operator is given by $H_2=\frac{1}{\sqrt 2}\bmatrix{1&1\\1&-1}.$ We refer to $H=H_2\otimes H_2$ as the Hadamard matrix, which operates as follows: 
  \beano H &: &\ket{0}\rightarrow (\ket{0}+\ket{1}+\ket{2}+\ket{3})/2, \\
 && \ket{1}\rightarrow (\ket{0}-\ket{1}+\ket{2}-\ket{3})/2,\\
  && \ket{2}\rightarrow (\ket{0}+\ket{1}-\ket{2}-\ket{3})/2,\\
   && \ket{3}\rightarrow (\ket{0}-\ket{1}-\ket{2}+\ket{3})/2.
 \eeano
We call a DTQW as the Hadamard walk when the coin operator is $H.$
 
Let $\ket{\Psi(t)}$ be the wave function at a given ``time'' $t\in \mathbb{N}.$ Then, we can write \beano \ket{\Psi(t)}=\sum_{(x,y)\in Z_t}\ket{x,y}\otimes\big(\psi_t^0(x,y)\ket{0}+\psi_t^1(x,y)\ket{1} \\
\phantom{bachchhara keu jhamela}+\psi_t^2(x,y)\ket{2}+\psi_t^3(x,y)\ket{3}\big),\eeano 
where $\psi^i_t(x,y)$ is the probability amplitude of the coin state $\ket{i}$ for $i=0,1,2,3,$ at position $(x,y)$ and time $t$. 
 With $t$ applications of the unitary $\U,$ the state of the walker and coin is $\ket{\Psi(t)}=\U^t\ket{\Psi(0)},$ where $\ket{\Psi(0)}=\ket{0,0}\otimes \ket{\phi_0}$ is the initial state, with the initial coin state $\ket{\phi_0}.$ For example, with the Grover coin operator, if $\ket{\phi_0}=(\ket{0}+\ket{1}-\ket{2}-\ket{3})/2,$ then 
 \beano \ket{\Psi(1)}&=&S(I\otimes C)\ket{0,0}\otimes \left(\ket{0}+\ket{1}-\ket{2}-\ket{3}\right)/2\\&=&(-\ket{1,0}\otimes \ket{0}-\ket{-1,0}\otimes \ket{1}+\ket{0,1}\otimes \ket{2}+\\&&
 \phantom{korona, ulTopalTa prashna}\ket{0,-1}\otimes \ket{3})/2.\eeano 
 
\begin{figure}[H] 
\centering
\includegraphics[height=6 cm,width=8 cm]{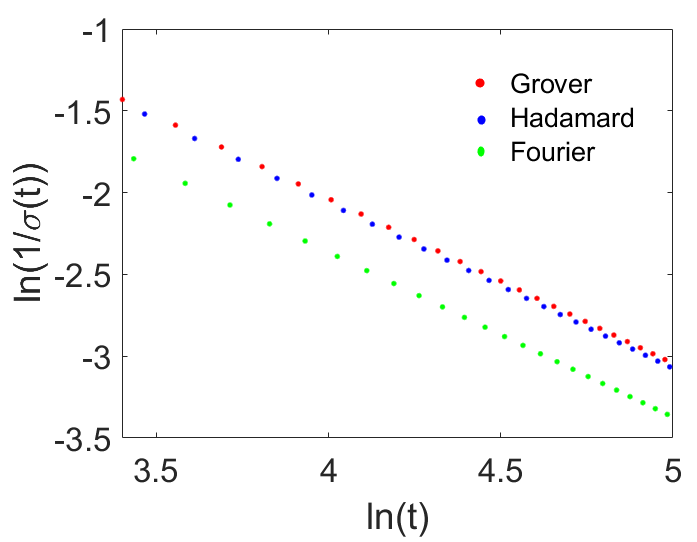}
\caption{ 
(Color online.) 
Ballistic spreads of clean 2D-DTQWs. A log-log plot illustrating the inverse of the standard deviation as a function of time is shown for 2D-DTQWs without disorder, using three different types of coins. Red, blue, and green dots correspond to Grover, Hadamard, and Fourier coins, respectively. In all the cases, we see a straight line fit with slope $-1$ satisfying $\ln(\frac{1}{\sigma(t)})=-\ln(t)+c,$ where $c$ changes for different coins. See text for the initial states of the coins. The walker starts at (0,0). Both axes represent dimensionless quantities.
}\label{fig:sd all}
\end{figure}

Let $$\rho(t)=\U^t\rho(0)(\U^\dagger)^t,$$ where $\rho(0)=\ket{\Psi(0)}\bra{\Psi(0)}.$
Let $$\overline{\rho}(t)=\mathrm{Tr}_c\rho(t),$$ where $\mathrm{Tr}_c$ denotes the partial trace taken over the coin degree of freedom.
The statistical properties of the quantum walker can be investigated via $P_t(x,y),$ the probability of finding the particle (walker) at position $(x,y)$ at time $t.$ Accordingly, the $k$th ($k=1,2$) statistical moment of the process at time $t$ is given by \cite{portugal2013quantum} $$m_k(t)=\sum_{(x,y)\in Z_t}(x^2+y^2)^{k/2}P_t(x,y),$$ where 
$$P_t(x,y)=\langle x,y|\overline{\rho}(t)|x,y\rangle=\sum_{j=0}^3|\psi_t^j(x,y)|^2.$$
It is well-known that for 1D DTQWs, $m_2(t) \sim t^{2},$ which results in $\sigma(t)=\sqrt{m_2(t)-m_1^2(t)} \sim t$, where by $f(t)\sim g(t),$ we mean that \(\frac{f(t)}{g(t)}\rightarrow\)
a finite value (non-zero and independent of \(t\)) in the limit $t \rightarrow \infty$~\cite{konno2002quantum, grimmett2004weak, konno2005new} 
Just like the one-dimensional case, two-dimensional DTQWs, with all the aforementioned coin operators, show ballistic spread \cite{portugal2013quantum}, i.e., $\sigma(t) \sim  t,$ and choice of the initial state vector does not affect the linear dependency of $\sigma(t)$ on $t.$ 
Let us illustrate this in Fig. \ref{fig:sd all}. We consider that the walker starts at $(0,0)$ and that the coin is initially in the state as indicated below:
\begin{eqnarray}
\label{harkin-choT}
\text{Grover}: && \frac{1}{2}\left(\ket{0}+\ket{1}-\ket{2}-\ket{3}\right),  
 \nonumber \\
\text{Fourier}: && \frac{1}{2\sqrt{2}}\left(\sqrt{2}(\ket{0}+\ket{2})+(1-i)(\ket{1} - \ket{3})\right), \nonumber \\  \text{Hadamard}: && \frac{1}{2}\left(\ket{0}+i\ket{1}-i\ket{2}+\ket{3}\right),
\end{eqnarray}
where the names on the left of each row points to the coin operation to be applied at each time step.
The ballistic spreads are clearly visible in all cases.

\section{Glassy disorder in DTQWs} \label{sec:DTQW disorder}
In this section, our focus is on the spread of 2D-DTQWs in the presence of ``glassy disorder''.
The term ``glassy disorder'' is used to indicate that the disorder incorporated in the dynamics will be such that its equilibration time will be several orders of magnitude higher than the times over which we observe the system. This mimics the glassy disorders considered in cooperative many-body phenomena~\cite{carleo2012localization,biroli2017delocalized, vojta2019disorder}.

The glassy disorder is incorporated here in the jump-length at every time step, and so instead of the  shifts of unit length in the four directions at each step of the 2D-DTQW, we assume a random jump-length that follows a certain probability distribution. See~\cite{revesz2013random} in this regard, for classical walks.
Let $J_t$ be the jump length at the $t$th time step, where $\{J_t\}_t$ are independent and identically distributed random variables according to a given discrete probability distribution $p(J_t)$. 
This type of disorder is termed as dynamic disorder, as the jump length at a certain time step is independent of that at all others.
 Here, the total state space equals $\mathrm{Span}\{\ket{x,y}\otimes\ket{{l}}|~(x,y)\in Z_{Rt},l\in\{0,1,2,3\}\}$ and is isomorphic to $ \C^{(2Rt+1)^2}\otimes\C^4,$ where $R\in \mathbb{Z_+}$ is the maximum jump length.
Thus, the conditional shift operator $S^d$ associated with this kind of disordered two-dimensional DTQW is as follows:
\begin{eqnarray*}
 S^d & = &\sum_{(x,y)\in Z_{R(t-1)}}\Big(\ket{x+J_t,y}\bra{x,y}\otimes \ket{0}\bra{0}+\\
 &&\phantom{phalgune tal j}\ket{x-J_t,y}\bra{x,y}\otimes \ket{1}\bra{1}+\\
 &&\phantom{phalgune tal j}\ket{x,y+J_t}\bra{x,y}\otimes \ket{2}\bra{2}+\\
 &&\phantom{phalgune tal j}\ket{x,y-J_t}\bra{x,y}\otimes \ket{3}\bra{3}\Big),
 \end{eqnarray*}
 where $J_0=0$, and $J_t\leq R$, for $t=1,2,\ldots$, are randomly chosen nonnegative integers from the given probability distribution $p(J_t)$.
 If the Grover walk starts with the initial coin state  $\ket{\phi_0}=(\ket{0}+\ket{1}-\ket{2}-\ket{3})/2$, then 
 \beano \ket{\Psi(1)}&=&S^d(I\otimes C)\ket{0,0}\otimes \left(\ket{0}+\ket{1}-\ket{2}-\ket{3}\right)/2\\&=&(-\ket{J_1,0}\otimes \ket{0}-\ket{-J_1,0}\otimes \ket{1}+\\&&\phantom{aa chal ke tu }\ket{0,J_1}\otimes \ket{2}+\ket{0,-J_1}\otimes \ket{3})/2.\eeano 
 One can similarly find the states in the next iterations, and in particular, in $\ket{\Psi(2)}$, the (unnormalized) position state conditioned to the coin basis state $\ket{0}$, i.e., \(_c\langle 0\ket{\Psi(2)}\), with the bra being of the dual Hilbert space of \(\mathcal{H}_c\), reads 
 $$(-\ket{J_2+J_1,0}-\ket{J_2-J_1,0}+\ket{J_2,J_1}-\ket{J_2,-J_1})/4.$$

We study the disorder-averaged dispersion of disordered 2D-DTQWs, with coins identified  by the operations they perform on the coin state at every time step, viz. \(G\), \(F\), and $H$, to measure the spread of the walker with time, by considering several random discrete distributions that generate different collections of $\{J_t\}_t$. Analyzing the scaling with  number of time steps of the disorder-averaged dispersion of disordered DTQWs helps in understanding the disorder strength and its relation with the spread of the walk. The strength of the disordered walk is related to the ``intensity'' of the randomness in the system, and  depends on the probability distribution selected for the random outputs $\{J_t\}_t$, and can be quantified by a measure of dispersion of the probability distribution. It is usual to use the standard deviation of the distribution as the measure of its dispersion.  The average value of dispersion of the disordered quantum walk measures the nature of the quantum walk in the presence of disorders. To quantify this, we evaluate the standard deviation,  $\sigma(t)$, for a sufficiently large set of realizations of the disordered jumps and consider the average, $\langle\sigma(t)\rangle$, at the $t$th time step. We compare the effects of different types of disorders on the walker's dynamics of 2D-DTQW by evaluating the corresponding $\langle\sigma(t)\rangle$ (see the details given in Tables~\ref{table:all} and~\ref{table:grov all}).

We now briefly describe various discrete probability distributions~\cite{ross2014first} that we utilize to generate random numbers that will act as  jump lengths in the disordered walks.

\subsection{Poissonian, sub-Poissonian, and super-Poissonian distributions}
The Poisson distribution, which is one of the most important discrete probability distributions, arises in many real-life situations. If $\lam$ is the mean number of events, then the probability mass function $p(k)$ of an event to occur for $k=0,1,2,\ldots$ times in a given time interval is, within Poisson distribution, given by $p(k)=\frac{\exp(-\lam)\lam^k}{k!}$. Clearly, if $\sigma^2$ is the variance, then $\frac{\sigma^2}{\lambda}=1$.  


Along with the Poisson distribution, we also consider other discrete probability distributions  to generate the random numbers for disordered walks, viz. the binomial, hypergeometric, negative binomial, and geometric distributions. Binomial and hypergeometric distributions are the sub-Poissonian distributions, having variances smaller than that of the Poisson distribution with an equal mean.
On the other hand, super-Poissonian distributions, like the negative binomial and geometric distributions, have variances satisfying the opposite inequality. \\

 \noindent \emph{Binomial distribution.} Let $N$ be the number of independent trials, each of which results in a success with probability $p.$ Then the probability mass function of a binomial random variable having parameters $(N,p)$ with $k$ successes for the particular sequence of $N$ outcomes is $p(k)={N \choose k} p^k (1-p)^{N-k},0\leq k \leq N.$ This is a sub-Poissonian type since we get $\frac{\sigma^2}{\lambda}=1-p<1.$ \\

\noindent \emph{Hypergeometric distribution.} 
Suppose a sample of size $n$ is chosen randomly from an urn containing $N$ elements as the population, of which $m$ is the exact number of elements termed as successes. Therefore, $N-m$ is the number of failures. If $k$ is the random variable of successes in $n$ draws from $N$ in a particular trial, then the probability mass function is $p(k)={m \choose k}{N-m \choose n-k} /{{N \choose n}}.$ This is a sub-Poissonian distribution with $\frac{\sigma^2}{\lambda}=\frac{N-k}{N}\frac{N-n}{N-1}<1.$\\ 

\noindent \emph{Negative binomial distribution.} Suppose that independent trials, each with a probability of success $p$, are performed until a total of $r$ successes occur. If $k$ equals the number of failures, then the probability mass function for the negative binomial distribution is $p(k)={k+r-1 \choose r-1} p^r (1-p)^{k}.$ Here $\frac{\sigma^2}{\lambda}=\frac{1}{p}>1,$ indicating the super-Poissonian nature of the distribution.\\

\noindent \emph{Geometric distribution.} In a geometric distribution, independent trials are performed until a success occurs. Suppose  
$p$ is the probability of success of each trial, and the random variable $k$ is the number of failures before the first success occurs. Then the probability mass function is $p(k)=(1-p)^kp$
with $\frac{\sigma^2}{\lambda}=\frac{1}{p}>1.$ Hence, this is a super-Poissonian distribution.\\

For all the  probability distributions decribed above, we generate the random numbers with a maximum length $R$  such that the probability of having random numbers greater than $R$ is  of order $10^{-4}$ or lower.

\section{Spreading analysis} \label{sec:analysis}
In this section, we analyze the spreading behavior of disordered 2D-DTQWs and their differences from the ordered 2D-DTQWs and the 2D-CRWs. We characterize the dynamics of the walks by computing the scaling exponent $ \alpha$, involved in the asymptotic relation $\sigma(t) \sim  t^{\alpha}$, where $\sigma(t)$ represents the standard deviation of the position probability distribution of the quantum walker. 

\subsection{Classical}
First, we consider the classical 2D random walks.
The probability distribution of a 1D-CRW is normal (also known as Gaussian) with $\sigma^c(t) \sim  t^\frac{1}{2},$ where $\sigma^c(t)$ denotes the standard deviation of the position probability distribution of the CR walker~\cite{chandrasekhar1943stochastic}. For a complete description of $\sigma^c(t)$ of a 2D-CRW, we first recall the following result from \cite{revesz2013random}.
\begin{theorem}\label{thm1}
Let $P^c_t(x,y)$ be the probability of the classical walker to be in position $(x,y)\in \mathbb{Z}^2$ after $t$ time steps. Then, 
\beano P^c_t(x,y)=\frac{1}{4^t} {t \choose \frac{t+x+y}{2}} {t \choose \frac{t+x-y}{2}},\eeano provided that $x+y=0(\text{mod}~2),|x|+|y|\leq t,$ and $t$ is even.
\end{theorem}
It can be checked that for $x+y=1(\mbox{mod}~2),|x|+|y|\leq t$, and $t$ odd, we get the same probability expression as given in Theorem~\ref{thm1}. Note that $P^c_t(x,y)=0$ in the other cases. 
 Using Stirling's approximation formula, $t!\sim t^t \exp(-t) \sqrt{2\pi t}$ as $t\rightarrow \infty,$ we simplify the relation from Theorem \ref{thm1} as
$$P^c_t(x,y)\sim \frac{2}{\pi t}\exp{\left(-\frac{x^2+y^2}{t}\right)},$$ so that for each fixed $t,$ $\frac{1}{2}P_t(x,y)$ is the bivariate normal distribution.
By converting the sum into an integral, the second moment, a measure used to describe the diffusion phenomena, can be expressed as
$$m^c_2(t)\sim \frac{1}{\pi t}\int_{-\infty}^{\infty}\int_{-\infty}^{\infty}{(x^2+y^2)}\exp{\left(-\frac{x^2+y^2}{t}\right)}\,dx\, dy.$$ The above expression takes the form 
$$m^c_2(t)\sim \frac{1}{\pi t}\int_{0}^{\infty}\int_{0}^{2\pi}r^3\exp{\left(-\frac{r^2}{t}\right)}\,dr\, d\theta,$$ in the polar coordinate system. Finally, the integration yields \(t \Gamma(2)\), so that $m^c_2(t)\sim t$. Similarly, the first moment gives $m^c_1(t)\sim t^\frac{1}{2}\frac{\sqrt \pi}{2}.$ Thus we get $\sigma^c(t)\sim t^\frac{1}{2}.$
Hence,  like the 1D case, position probability distributions of 2D CRWs follow Gaussian distributions and $\sigma^c(t) \sim t^{\frac{1}{2}}$. It is worth mentioning that for Poisson-disordered CRWs, where the jump lengths stem from Poisson distributions  
at each step, the disordered-averaged standard deviation of the position distribution  asymptotically behave as  
$t^{\frac{1}{2}}$ with time step, $t$, on both 1D and 2D lattices. 
Hence, there are no significant changes in the spreading behavior of 1D or 2D CRWs, with the insertion or onset of disorder.
We demonstrate this by 
simulating the disordered 2D-CRW, where the jump lengths are Poisson-distributed with a certain mean, as discussed 
in 
Section~\ref{sec:DTQW disorder} for disordered 2D-DTQWs. 
We can use the symmetric discrete-time iterative map,
\beano P^c_{t+1}(x,y) &=& \frac{1}{4}P^c_t(x - J_t,y) + \frac{1}{4} P^c_t(x+J_t,y)+\\&&\frac{1}{4}P^c_t(x,y - J_t) + \frac{1}{4} P^c_t(x,y+J_t),\eeano  where $J_t$ is a randomly chosen Poisson-distributed integer at the $t$th step. 
See~\cite{kalikow1981poisson} in this regard.

\subsection{Quantum}
The celebrated result concerning DTQWs on a line with $H_2$ as the coin operator states that $\sigma(t)$ of the corresponding position probability distribution 
can be approximated by 
 $\sqrt{1-\frac{1}{\sqrt 2}}t$,
for a large number of time steps, \(t\)
\cite{konno2002quantum}.
It was later on found that 
the linear relation of $\sigma(t)$ with $t$ changes significantly 
in 
presence of disorder in the conditional shift operator, and the corresponding walk becomes sub-ballistic but remains super-diffusive, with the scaling exponent of $t$ lying between one-half and unity~\cite{das2019inhibition}.

 We now analyze the dynamics of 2D-DTQWs in presence of  disorder in the jump length for several paradigmatic distributions of disorder by calculating the corresponding scaling exponents of the spread in position, i.e., the $\alpha$ of  standard deviation, $\sigma(t) \sim t^ \alpha$, of the corresponding position distribution, where \(t\) is the number of time steps, and where the scaling is considered for large \(t\). 
 The time-scaled long-time limit distribution of the Grover walk - in absence of disorder - indicates a ballistic spreading (\(\alpha = 1\))~\cite{watabe2008limit}. The same holds for \(H\) and \(F\) as coin operations also - again in absence of disorder.
 See Fig.~\ref{fig:sd all}. So just like in 1D, the scaling exponent of 2D-DTQWs is also 
 double that of the 2D-CRWs, in absence of disorder. In all the instances, we assume that the walker starts at $(0,0)$, and that the initial coin state is as in Eq.~(\ref{harkin-choT}). 
 The choice of the initial coin states 
 result in  symmetric position probability distributions about the lines $y=x$ and $y=-x$ through the origin in the $x\,y$\nobreakdash-plane, which is clear from panels~$(a)$ and the corresponding contour plots in panels~$(b)$ of Figs.~\ref{fig:2d Grover},~\ref{fig:2d Fourier}, and~\ref{fig:2d Hadamard}.
Meanwhile, with the same set of initial states, the presence of disorder leads to a significantly different set of patterns for the position distribution. 
The position probability distributions of disordered DTQWs are recorded, for  given realizations of the disorder, in Figs.~\ref{fig:2d Grover}(c),~\ref{fig:2d Fourier}(c), and~\ref{fig:2d Hadamard}(c), and corresponding 2D contour diagrams are given in Figs.~\ref{fig:2d Grover}(d),~\ref{fig:2d Fourier}(d), and~\ref{fig:2d Hadamard}(d). 

\begin{widetext}

\begin{figure}[ht] 
\centering
\subfigure[ ]{\includegraphics[height=4 cm,width=6 cm]{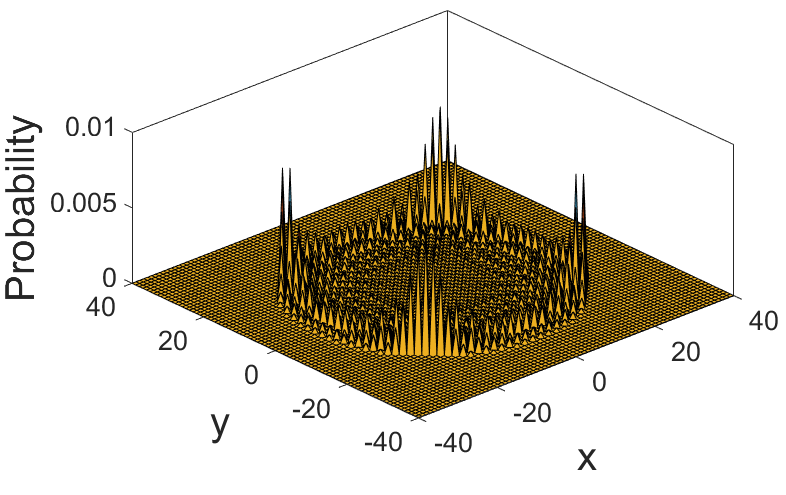}}
\subfigure[ ]{\includegraphics[height=4 cm,width=6 cm]{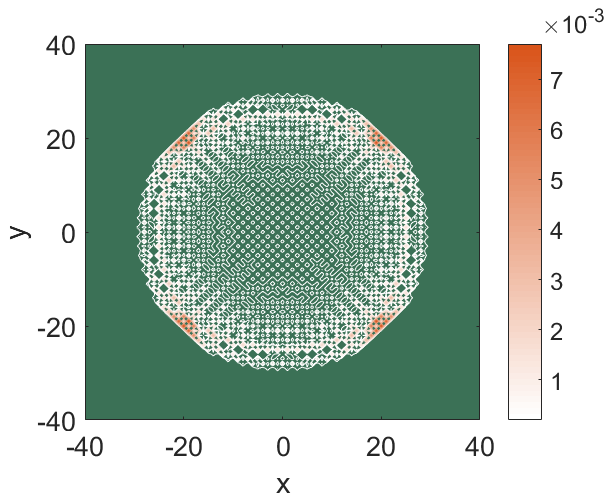}}\\
\subfigure[]{\includegraphics[height=4 cm,width=6 cm]{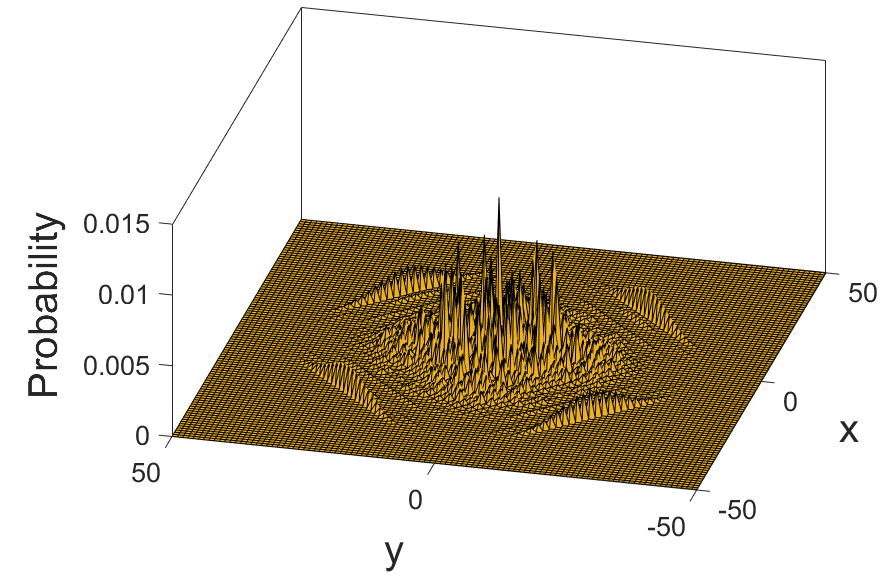}} 
\subfigure[ ]{\includegraphics[height=4 cm,width=6 cm]{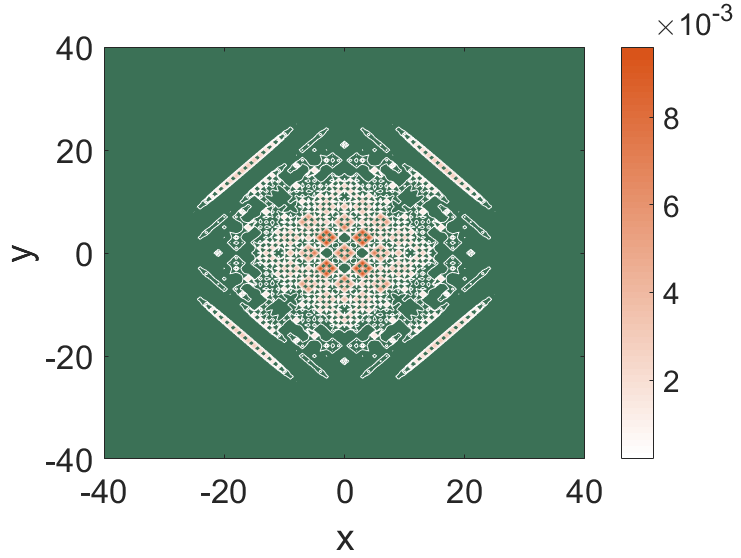}}
\caption{Inhibition of spread of Grover walk due to insertion of disorder in the jump length of the 2D-DTQW. Panels~(a) and~(c) exhibit the position probability distributions of Grover walks after $t=40$ time steps without and with disorder, respectively, where the initial coin state vector is 
\([1/2,1/2,-1/2,-1/2]^T\), in the computational basis. Correspond to (a) and (c), the 2D contour plots of the probabilities
are given in panels~(b) and~(d), respectively. The two directions on the 2D plane on which the walker is moving are represented on the \((x,y)\) plane. The disorder inflicted is a particular realization of the sequence of jump lengths, chosen independently from the Poisson distribution with unit mean. All quantities used in the plots are dimensionless.}\label{fig:2d Grover}
\end{figure} 

\begin{figure}[ht] 
\centering 
\subfigure[ ]{\includegraphics[height=4 cm,width=6 cm]{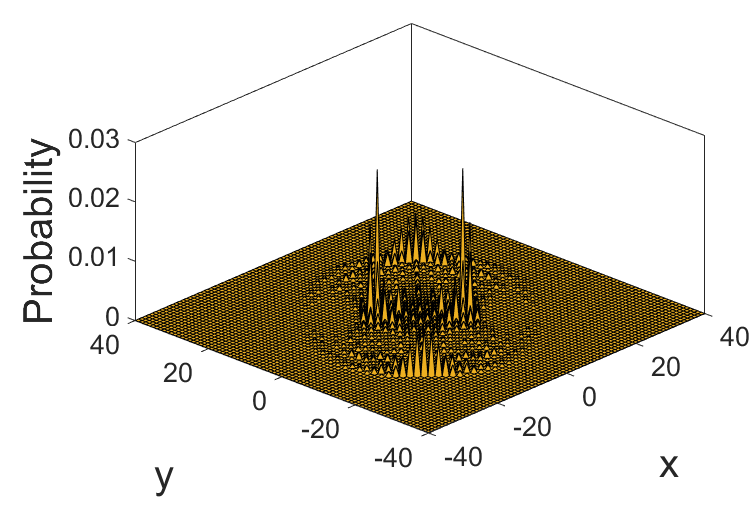}}
\subfigure[ ]{\includegraphics[height=4 cm,width=6 cm]{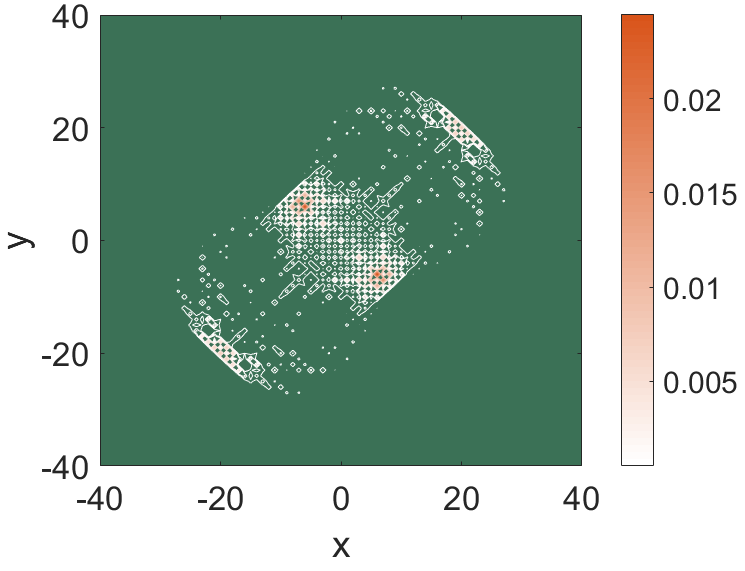}}
\subfigure[]{\includegraphics[height=4 cm,width=6 cm]{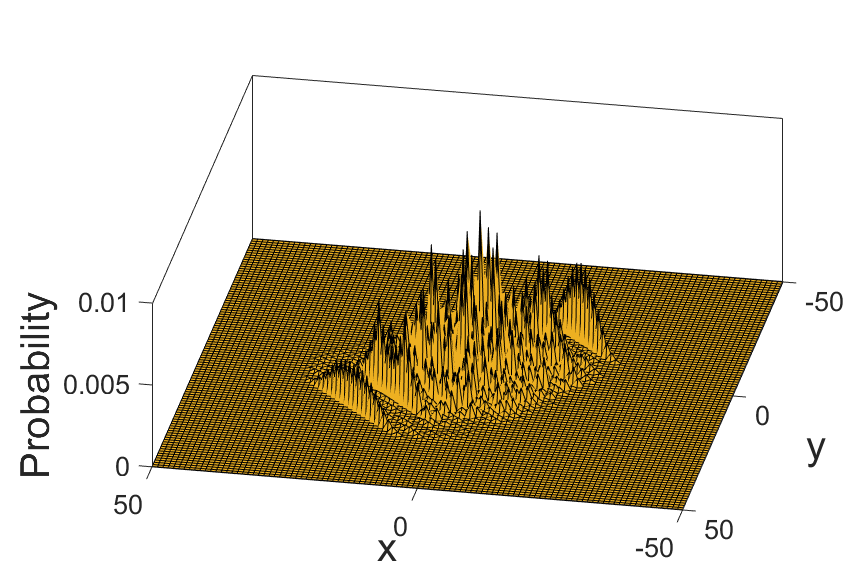}}
\subfigure[ ]{\includegraphics[height=4 cm,width=6 cm]{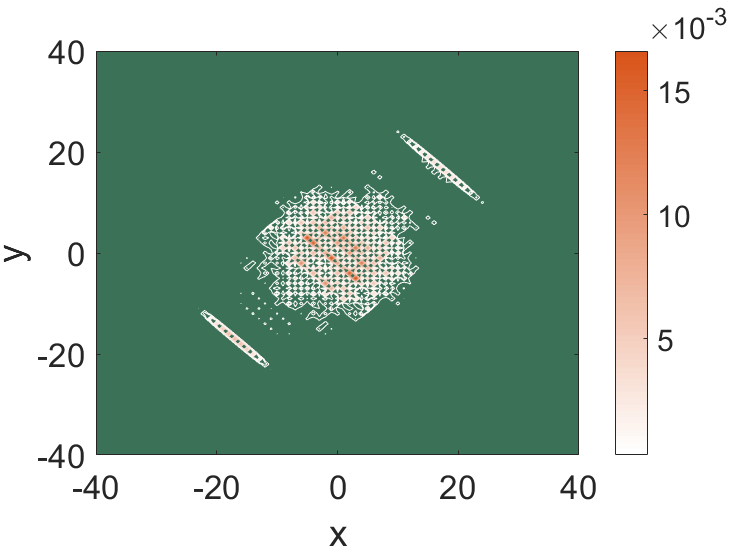}}
\caption{Inhibition of spread due to disorder in jump length for Fourier walk on 2D square lattice. The considerations here are exactly the same as in the preceding figure, except that the initial coin state here is 
\(\frac{1}{2}[1,(1-i)/\sqrt{2},1,-(1-i)/\sqrt{2}]^T\), and the coin operator is the Fourier transformation. All quantities plotted are dimensionless.
}\label{fig:2d Fourier}
\end{figure} 

\begin{figure}[ht] 
\centering
\subfigure[ ]{\includegraphics[height=4 cm,width=6 cm]{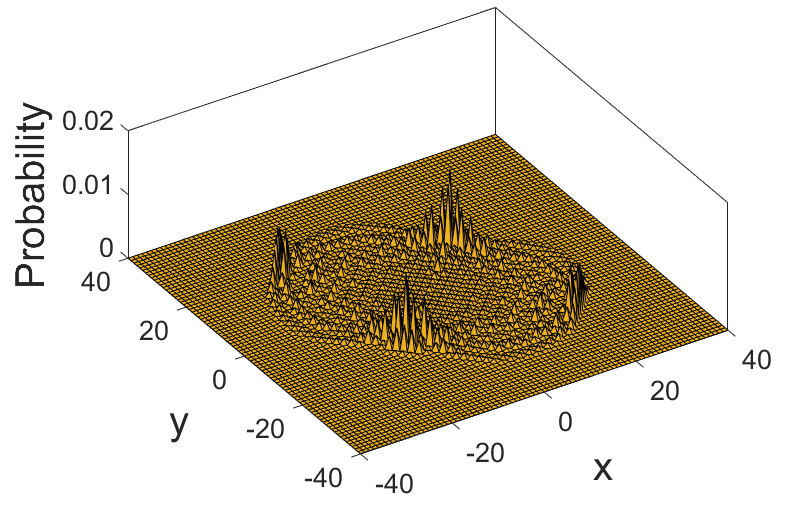}}
\subfigure[ ]{\includegraphics[height=4 cm,width=6 cm]{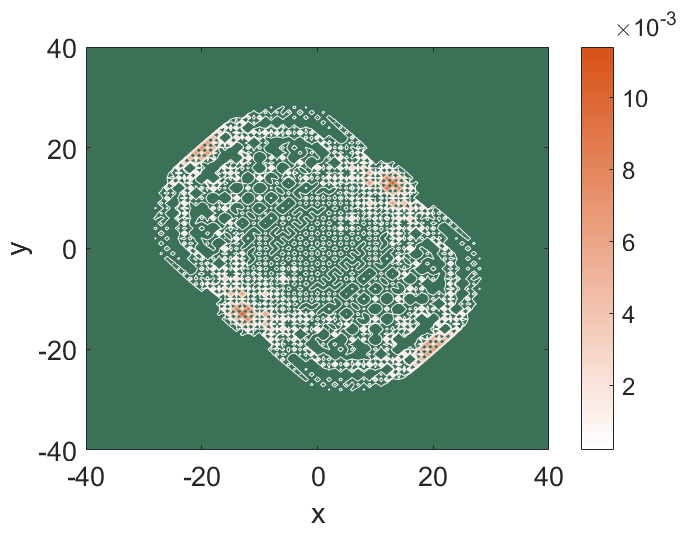}}
\subfigure[]{\includegraphics[height=4 cm,width=6 cm]{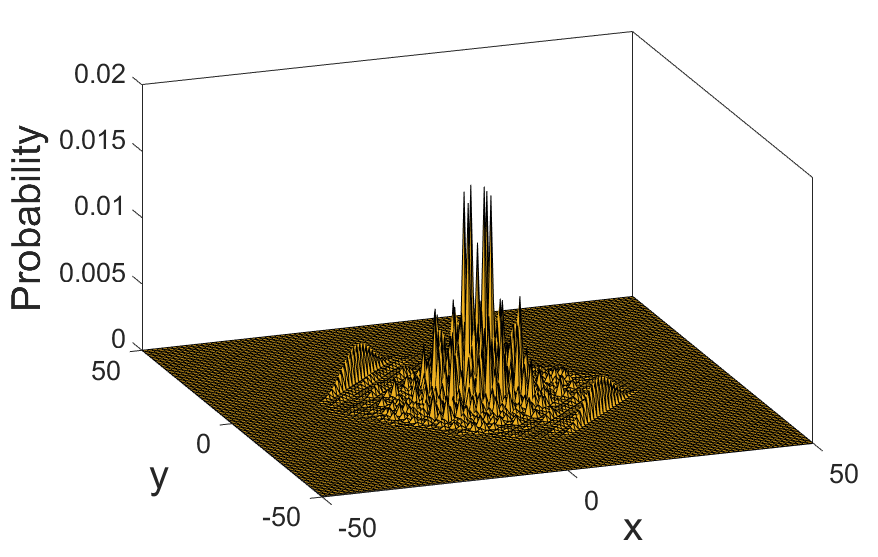}} \nonumber
\subfigure[ ]{\includegraphics[height=4 cm,width=6 cm]{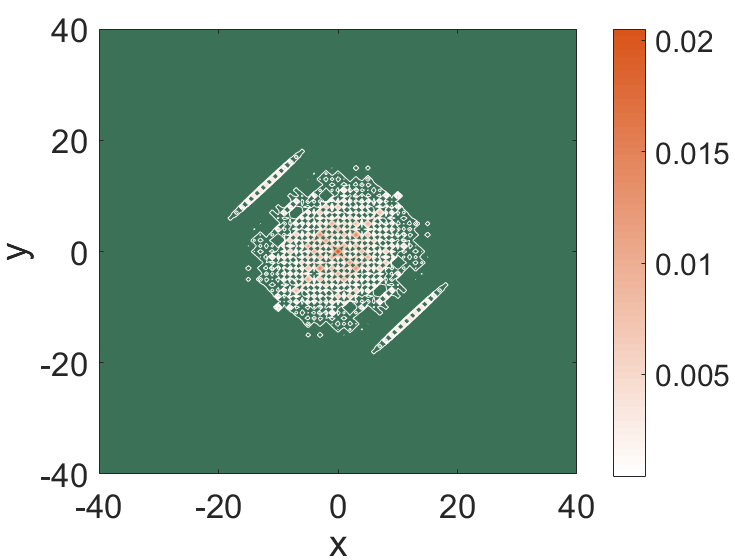}}
\caption{Inhibition of spread due to disorder insertion in jump length for Hadamard walk on 2D square lattice. The considerations here are again exactly the same as in Fig.~\ref{fig:2d Grover}, except that the initial coin state here is
\([1/2,i/2,-i/2,1/2]^T\), and the coin operator is \(H\). All quantities plotted are dimensionless.
}\label{fig:2d Hadamard}
\end{figure} 
\end{widetext}

Panels~(a) and~(c) in Figs.~\ref{fig:2d Grover},~\ref{fig:2d Fourier}, and~\ref{fig:2d Hadamard}, compare the position probabilities of the 2D-DTQWs without and with disorders after $t=40$ time steps, where the walks are defined by the Grover, Fourier, and Hadamard coins. In each scenario, the walker starts from $(0,0)$ and propagates to the other vertices of the lattice as time progresses. Panels (a) exhibit the probability distribution for the ordered or clean walks. 
Panels (c) show probabilities for the disordered walks for a single realization of the Poisson-distributed disorder with a mean value $1$. Disorders in 2D-DTQW inhibit spreading of the walk through the lattice, and the walker remains closer to the origin in comparison to the case when there is no disorder. This feature of disordered walks can be easily understood from the contour diagrams in panels~(d) of Figs.~\ref{fig:2d Grover},~\ref{fig:2d Fourier}, and~\ref{fig:2d Hadamard}. We observe this special characteristic of the disordered 2D-DTQWs not only for this particular one but also for other realizations of the disorder probability distribution. In the presence of disorder, taking the average over a sufficiently large number of disorder realizations helps to understand this spreading dynamic and transport phenomena clearly - and independently of the particular realization of the disorder.\\

A comment about the choice of the initial coin state is in order here. These are chosen so that the corresponding position distributions are as symmetric as possible around the initial walker position. However, we have checked for several other instances of the initial coin state that this choice does not affect the scaling exponents of the walkers' spread in position.\\

\begin{figure}[H] 
\centering
\includegraphics[height=6 cm,width=7 cm]{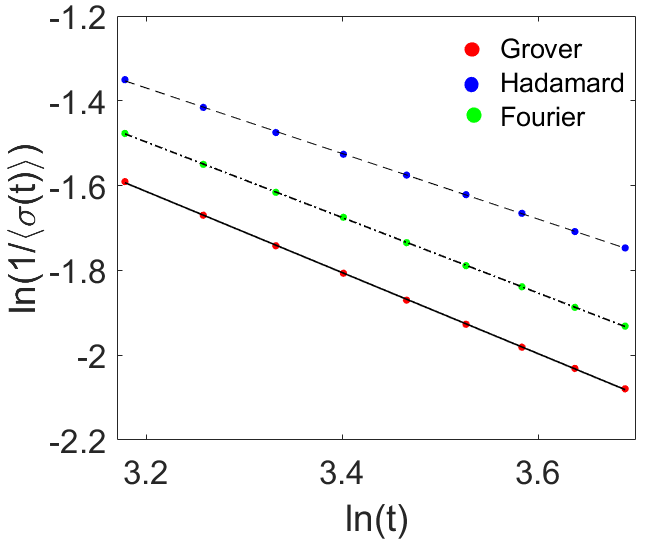}
\caption{(Color online.) 
Disorder-induced inhibition of spread of quantum walk in 2D. The plots shows the extent to which a Poisson-disordered jump length with unit mean inhibits the spread of the position distribution of a quantum walker on a 2D square lattice. The
red, blue, and green colored dots correspond to quantum walks with the coin operators, \(G\), \(H\), and $F$, respectively. 
The linear fittings (on the log-log plot) for the data points (colored dots) corresponding to the Grover, Hadamard, and Fourier walks are represented by solid, dashed, and dash-dotted black lines, respectively. With $95\%$ confidence level, the slopes of the fitted lines for $G, H,$ and $F$ coins fall within the confidence intervals 
$-0.96\pm 0.005, -0.77\pm 0.004$ and $0.89\pm 0.004,$
respectively, with an average least square error of $0.002$ for the linear fittings.  
For the purpose of the fittings, we have focused on the range \(17 < t < 50\). 
}\label{fig:sd all disorder}
\end{figure}

\noindent \textbf{Scaling for Poisson disorder with unit mean.} For disordered DTQWs, we  numerically evaluate $\langle \sigma(t)\rangle$, where $t$ varies from $18$ to $50$, for a variety of  discrete probability distributions  with different strengths. The notation \(\langle \cdot \rangle\) denotes an average of the argument over the corresponding disorder realizations. We begin with the Poisson distribution with unit mean, and in Fig.~\ref{fig:sd all disorder}, we depict the dependency of $\langle \sigma(t) \rangle$ with time $t$ in disordered 2D-DTQWs, for $G, F,$ and $H$ coins.
We plot the inverse of $\langle \sigma(t)\rangle$ along the vertical direction against changes in $t$ along the horizontal axis. For a clear visualization of the relationship in the data, we consider log-log (natural log) 
plots. All numbers
are correct up to two
significant figures. We run the process to evaluate $\langle \sigma(t)\rangle$ for different sets of disorder realizations with increasing size until the fitted scaling values in the log-log plots of $1/\langle \sigma(t)\rangle$ and  $t$ for two consecutive runs are the same up to two significant figures. 
For the final set of disorder realization, we record the fitted scaling exponent and the corresponding confidence interval with $95\%$ confidence level. This method for recording the final scaling exponent continues on other occasions throughout the paper.
We find that the scalings of the disorder-averaged standard deviations can have significantly different values depending on the coin operator, with the Grover coin showing maximum resistance to inhibition of spread, among the coins considered. However, the scaling remains sub-ballistic but super-diffusive (i.e., between \(1/2\) and \(1\)) for all coin operators considered. 
\\


\noindent \textbf{Effect  of different means of the Poisson-disordered jumps on scaling of spread.} In a clean 2D-DTQW, the walker moves ``linearly'' with time, i.e., the standard deviation of the position probability distribution scales linearly with the number of steps. We have seen that the probability distribution changes significantly when a disorder is inserted in the jump length of the walker, with the disorder being Poisson distributed with unit mean. 
It is interesting to find out if the mean of the disorder has a significant effect on the scaling of the spread. We tabulate the scalings of the disorder-averaged standard deviations with jump lengths being chosen from Poisson distributions of  different means. See Table~\ref{table:all}. We find that the response to change in the jump length can be as large as a change in the coin operator. However, again the scaling remains sub-ballistic but super-diffusive for all coins. And again the Grover coin provides the maximum resilience to disorder-induced inhibition of spread of the walker among the coins considered.  
\\

\begin{table}[ht]
\centering
\begin{tabular}{c c c c} 
 \hline
Coin & Distribution & \quad $\lambda$ \quad \quad & $\alpha$\\ [0.5ex] 
 \hline
\multirow{3}{4em}{Grover} & Poisson & 0.7 & 0.99  \\ 
 & Poisson & 1 & 0.96 \\ 
 & Poisson & 1.4 & 0.93 \\
 & Poisson & 2 & 0.92  \\ [1ex] 
 \hline
 \multirow{3}{4em}{Fourier} &  Poisson & 0.7 & 0.93 \\ 
& Poisson & 1 & 0.89  \\
 & Poisson & 1.4 & 0.86 \\
 & Poisson & 2 & 0.84  \\ [1ex] 
 \hline
 \multirow{3}{4em}{Hadamard} &  Poisson & 0.7 & 0.78 \\ 
& Poisson & 1 & 0.77  \\
 & Poisson & 1.4 & 0.74 \\
 & Poisson & 2 & 0.72  \\ [1ex] 
 \hline
\end{tabular}
\caption{
Response of scaling of spread to change in mean of disorder. While remaining within the domain of Poisson-disordered jump lengths, we investigate here the effect of a variation in the mean of the disorder distribution on the scaling (\(\alpha\)) of the disorder-averaged standard deviation of the position probability distribution of the 2D quantum walker, for Grover, Fourier, and Hadamard coin operators. 
We find that there is generally an increase in the scaling exponent with increase in the mean (\(\lambda\)) of the disorder distribution. All quantities in the table are dimensionless. In each case, the 
true
value of $\alpha$ falls in the confidence interval with $95\%$ certainty, and the average length of the confidence interval is $0.008$.
}\label{table:all}
\end{table}

\noindent \textbf{Effect  of sub- and super-Poisson -disordered jumps on scaling of spread.} We next try to find out how the scaling exponent of the spread in position distribution of the disordered quantum walk is affected by altering the disorder distribution of the jump length of the walker. We check for incorporation of both sub- and super-Poissonian distributions. In Table~\ref{table:grov all}, we put the data for the scaling exponent of disordered Grover walks with disorder lengths chosen from several sub-Poissonian and super-Poissonian probability distributions, all with unit mean, and compare them with the Poissonian Grover walk with unit mean. There is a sub-ballistic but super-diffusive nature of the walker in each case. 

\begin{table}[H]
\centering
\begin{tabular}{c c c c} 
 \hline
Coin & Distribution & Variance & $\alpha$\\ [0.5ex] 
 \hline
\multirow{5}{4em}{Grover} & Poisson & 1 & 0.96 \\ 
& Hypergeometric & 0.6 & 0.99 \\
& Binomial & 0.8 & 0.97 \\
 & Geometric & 2 & 0.91 \\
 & Negative binomial & 2 & 0.93  \\ [1ex] 
 \hline
\end{tabular} 
\caption{Response of scaling of spread to change in disorder distribution. We focus on the Grover quantum walk in 2D and on disorder in the jump length with distributions of unit mean, but veer the distribution away from the Poissonian one. For the analysis, we consider two sub-Poisson and two super-Poisson distributions. A sub-Poisson distribution is one which has a lower variance than the Poisson distribution with the same mean. A super-Poissonian distribution has higher variance than the same. We remember that a Poisson distribution has equal mean and variance. The table exhibits the variances of the distributions and the corresponding scaling exponents (\(\alpha\)).
%
In each case, the 
true
value of $\alpha$ falls within the corresponding confidence interval with $95\%$ certainty, with the average length of the confidence interval being \(0.008\).
}   \label{table:grov all}
\end{table}

\subsection{Comparison with 1D discrete-time quantum walks}

In disordered 1D-DTQWs with the (single-qubit) Hadamard coin \(H_2\), the scaling exponent $\alpha$ is 
$0.75\pm 0.005$
with $95\%$ confidence level, when a Poisson distribution with unit mean is utilized to mimic the disorder in the jump length.
In the 2D scenario, for  Grover and Fourier walks, the  value of $\alpha$ is often close $0.9$, which is significantly larger than the 1D case. Indeed,  the Grover walk with jump-length disorder distributed as Poisson with mean 0.7, has a scaling exponent very close to unity, which is the exponent for the clean case in 2D (as well as in 1D). See Table~\ref{table:all}. 
This stronger resilience in 2D Grover and Fourier walks to disorder persists for variation in the mean of the Poisson distribution, and also for variation in the distribution itself. Therefore, the ``quantum  advantage'' of quantum walks over classical ones remains more secure in 2D than in 1D. It is therefore plausible that choosing 2D quantum walks over 1D ones will be more reasonable when utilizing the faster spread of quantum walks in a quantum technology.

\section{Static disorder} \label{sec:static}
In this section, we briefly discuss the role of static disorder in the spread of the 2D-DTQWs. Here, the jump length at each vertex is independent of time but random with respect to the vertex positions. At a vertex, the jump of the walker for going to the next vertex is decided according to the random output prescribed for that vertex according to a given probability distribution.
We perform numerical simulations for the Grover walk with static disorders due to Poissonian, sub-Poissonian, and super-Poissonian distributions. We choose all the probability distributions with unit mean  to evaluate the random integers before each realization of the process, and that is kept  unaltered at the vertices for the whole time of the evaluation. We generate numerous sets of such random numbers realizations and obtain the disorder-averaged value of standard deviation for $t\leq 50$. The results obtained are similar to those in 
Table~\ref{table:grov all}. This suggests that both static and dynamic 
disorders affect the Grover 2D walk similarly, at least for the 
spreading rate.


\section{Conclusion} \label{sec:conclusion}
In this paper, we consider discrete-time quantum walks on the two-dimensional square lattice with three types of coins: Grover, Fourier, and Hadamard. A classical walker can travel up to a distance $\sqrt{t}$ from the origin (starting point) after $t$ time steps, in the sense that the standard deviation of the walker's position distribution scales as so.  In contrast, the quantum walker reaches the distance $t$ at the same time, in the same sense. In other words, there is a quadratic speedup in DTQWs over  CRWs, defined on two-dimensional lattices. The same speedup happens for 
walks on 1D lattices. 

Some problems, like searching marked vertices on graphs employing quantum algorithms,  use quantum walks as key tools to speed up the process over the corresponding classical methods. 
Thus, analyzing the scaling exponent of the position probability distribution
of quantum walks
is crucial for understanding the efficiency of the underlying algorithm.
But in real scenarios, some unavoidable interactions between the quantum system and the environment, as well as some impurities, influence the system by inducing disorders, which motivated us to quantify the effect on the propagation rate of quantum walks due to the presence of noise. We showed that the existence of glassy disorder in the shift operation - due to a glassy disorder in the jump length - of 2D-DTQWs affects the walker's propagation by hindering the ``velocity''. The spreading is sub-ballistic and super-diffusive, i.e., the exponent on the number of time-steps of the disorder-averaged standard deviation of the position distribution lies between one-half and unity. 

We primarily considered Poisson distributions to induce random irregularities in the step lengths of the quantum walker at each step. Subsequently, we also studied the effects of sub- and super-Poissonian probability distributions. In all such cases, for disordered 2D-DTQWs with different coins, we observed that the ballistic spread of the clean quantum walker is impeded, but it remains significantly more quantum when $G$ and $F$ are the underlying coins, in comparison to the disordered 1D-DTQWs (with \(H_2\) as the coin operator). 
This indicates that it may be better for a quantum technology, that intends to use the faster spread of a quantum walker (than the corresponding classical walker) to its advantage, to employ two-dimensional quantum walks than one-dimensional ones, ignoring  the possibly higher amounts of resource necessary to build the former.




\begin{acknowledgments} We thank the cluster facilities at the Harish-Chandra Research Institute.  We acknowledge partial support from the Department of Science and Technology, Government of India through the QuEST grant (grant number DST/ICPS/QUST/Theme-3/2019/120). \end{acknowledgments}





\begin{thebibliography}{73}%
\makeatletter
\providecommand \@ifxundefined [1]{%
 \@ifx{#1\undefined}
}%
\providecommand \@ifnum [1]{%
 \ifnum #1\expandafter \@firstoftwo
 \else \expandafter \@secondoftwo
 \fi
}%
\providecommand \@ifx [1]{%
 \ifx #1\expandafter \@firstoftwo
 \else \expandafter \@secondoftwo
 \fi
}%
\providecommand \natexlab [1]{#1}%
\providecommand \enquote  [1]{``#1''}%
\providecommand \bibnamefont  [1]{#1}%
\providecommand \bibfnamefont [1]{#1}%
\providecommand \citenamefont [1]{#1}%
\providecommand \href@noop [0]{\@secondoftwo}%
\providecommand \href [0]{\begingroup \@sanitize@url \@href}%
\providecommand \@href[1]{\@@startlink{#1}\@@href}%
\providecommand \@@href[1]{\endgroup#1\@@endlink}%
\providecommand \@sanitize@url [0]{\catcode `\\12\catcode `\$12\catcode
  `\&12\catcode `\#12\catcode `\^12\catcode `\_12\catcode `\%12\relax}%
\providecommand \@@startlink[1]{}%
\providecommand \@@endlink[0]{}%
\providecommand \url  [0]{\begingroup\@sanitize@url \@url }%
\providecommand \@url [1]{\endgroup\@href {#1}{\urlprefix }}%
\providecommand \urlprefix  [0]{URL }%
\providecommand \Eprint [0]{\href }%
\providecommand \doibase [0]{https://doi.org/}%
\providecommand \selectlanguage [0]{\@gobble}%
\providecommand \bibinfo  [0]{\@secondoftwo}%
\providecommand \bibfield  [0]{\@secondoftwo}%
\providecommand \translation [1]{[#1]}%
\providecommand \BibitemOpen [0]{}%
\providecommand \bibitemStop [0]{}%
\providecommand \bibitemNoStop [0]{.\EOS\space}%
\providecommand \EOS [0]{\spacefactor3000\relax}%
\providecommand \BibitemShut  [1]{\csname bibitem#1\endcsname}%
\let\auto@bib@innerbib\@empty
\bibitem [{\citenamefont {Childs}(2009)}]{childs2009universal}%
  \BibitemOpen
  \bibfield  {author} {\bibinfo {author} {\bibfnamefont {A.~M.}\ \bibnamefont
  {Childs}},\ }\bibfield  {title} {\bibinfo {title} {Universal computation by
  quantum walk},\ }\href@noop {} {\bibfield  {journal} {\bibinfo  {journal}
  {Physical {R}eview {L}etters}\ }\textbf {\bibinfo {volume} {102}},\ \bibinfo
  {pages} {180501} (\bibinfo {year} {2009})}\BibitemShut {NoStop}%
\bibitem [{\citenamefont {Childs}\ \emph {et~al.}(2013)\citenamefont {Childs},
  \citenamefont {Gosset},\ and\ \citenamefont {Webb}}]{childs2013universal}%
  \BibitemOpen
  \bibfield  {author} {\bibinfo {author} {\bibfnamefont {A.~M.}\ \bibnamefont
  {Childs}}, \bibinfo {author} {\bibfnamefont {D.}~\bibnamefont {Gosset}},\
  and\ \bibinfo {author} {\bibfnamefont {Z.}~\bibnamefont {Webb}},\ }\bibfield
  {title} {\bibinfo {title} {Universal computation by multiparticle quantum
  walk},\ }\href@noop {} {\bibfield  {journal} {\bibinfo  {journal} {Science}\
  }\textbf {\bibinfo {volume} {339}},\ \bibinfo {pages} {791} (\bibinfo {year}
  {2013})}\BibitemShut {NoStop}%
\bibitem [{\citenamefont {Ambainis}(2003)}]{ambainis2003quantum}%
  \BibitemOpen
  \bibfield  {author} {\bibinfo {author} {\bibfnamefont {A.}~\bibnamefont
  {Ambainis}},\ }\bibfield  {title} {\bibinfo {title} {Quantum walks and their
  algorithmic applications},\ }\href@noop {} {\bibfield  {journal} {\bibinfo
  {journal} {International Journal of Quantum Information}\ }\textbf {\bibinfo
  {volume} {1}},\ \bibinfo {pages} {507} (\bibinfo {year} {2003})}\BibitemShut
  {NoStop}%
\bibitem [{\citenamefont {Childs}\ \emph {et~al.}(2003)\citenamefont {Childs},
  \citenamefont {Cleve}, \citenamefont {Deotto}, \citenamefont {Farhi},
  \citenamefont {Gutmann},\ and\ \citenamefont
  {Spielman}}]{childs2003exponential}%
  \BibitemOpen
  \bibfield  {author} {\bibinfo {author} {\bibfnamefont {A.~M.}\ \bibnamefont
  {Childs}}, \bibinfo {author} {\bibfnamefont {R.}~\bibnamefont {Cleve}},
  \bibinfo {author} {\bibfnamefont {E.}~\bibnamefont {Deotto}}, \bibinfo
  {author} {\bibfnamefont {E.}~\bibnamefont {Farhi}}, \bibinfo {author}
  {\bibfnamefont {S.}~\bibnamefont {Gutmann}},\ and\ \bibinfo {author}
  {\bibfnamefont {D.~A.}\ \bibnamefont {Spielman}},\ }\bibfield  {title}
  {\bibinfo {title} {Exponential algorithmic speedup by a quantum walk},\
  }\href@noop {} {\bibfield  {journal} {\bibinfo  {journal} {Proceedings of the
  thirty-fifth annual ACM symposium on Theory of computing}\ ,\ \bibinfo
  {pages} {59}} (\bibinfo {year} {2003})}\BibitemShut {NoStop}%
\bibitem [{\citenamefont {Shenvi}\ \emph {et~al.}(2003)\citenamefont {Shenvi},
  \citenamefont {Kempe},\ and\ \citenamefont {Whaley}}]{shenvi2003quantum}%
  \BibitemOpen
  \bibfield  {author} {\bibinfo {author} {\bibfnamefont {N.}~\bibnamefont
  {Shenvi}}, \bibinfo {author} {\bibfnamefont {J.}~\bibnamefont {Kempe}},\ and\
  \bibinfo {author} {\bibfnamefont {K.~B.}\ \bibnamefont {Whaley}},\ }\bibfield
   {title} {\bibinfo {title} {Quantum random-walk search algorithm},\
  }\href@noop {} {\bibfield  {journal} {\bibinfo  {journal} {Physical Review
  A}\ }\textbf {\bibinfo {volume} {67}},\ \bibinfo {pages} {052307} (\bibinfo
  {year} {2003})}\BibitemShut {NoStop}%
\bibitem [{\citenamefont {Ambainis}(2007)}]{ambainis2007quantum}%
  \BibitemOpen
  \bibfield  {author} {\bibinfo {author} {\bibfnamefont {A.}~\bibnamefont
  {Ambainis}},\ }\bibfield  {title} {\bibinfo {title} {Quantum walk algorithm
  for element distinctness},\ }\href@noop {} {\bibfield  {journal} {\bibinfo
  {journal} {SIAM Journal on Computing}\ }\textbf {\bibinfo {volume} {37}},\
  \bibinfo {pages} {210} (\bibinfo {year} {2007})}\BibitemShut {NoStop}%
\bibitem [{\citenamefont {Oka}\ \emph {et~al.}(2005)\citenamefont {Oka},
  \citenamefont {Konno}, \citenamefont {Arita},\ and\ \citenamefont
  {Aoki}}]{oka2005breakdown}%
  \BibitemOpen
  \bibfield  {author} {\bibinfo {author} {\bibfnamefont {T.}~\bibnamefont
  {Oka}}, \bibinfo {author} {\bibfnamefont {N.}~\bibnamefont {Konno}}, \bibinfo
  {author} {\bibfnamefont {R.}~\bibnamefont {Arita}},\ and\ \bibinfo {author}
  {\bibfnamefont {H.}~\bibnamefont {Aoki}},\ }\bibfield  {title} {\bibinfo
  {title} {Breakdown of an electric-field driven system: a mapping to a quantum
  walk},\ }\href@noop {} {\bibfield  {journal} {\bibinfo  {journal} {Physical
  Review Letters}\ }\textbf {\bibinfo {volume} {94}},\ \bibinfo {pages}
  {100602} (\bibinfo {year} {2005})}\BibitemShut {NoStop}%
\bibitem [{\citenamefont {Mohseni}\ \emph {et~al.}(2008)\citenamefont
  {Mohseni}, \citenamefont {Rebentrost}, \citenamefont {Lloyd},\ and\
  \citenamefont {Aspuru-Guzik}}]{mohseni2008environment}%
  \BibitemOpen
  \bibfield  {author} {\bibinfo {author} {\bibfnamefont {M.}~\bibnamefont
  {Mohseni}}, \bibinfo {author} {\bibfnamefont {P.}~\bibnamefont {Rebentrost}},
  \bibinfo {author} {\bibfnamefont {S.}~\bibnamefont {Lloyd}},\ and\ \bibinfo
  {author} {\bibfnamefont {A.}~\bibnamefont {Aspuru-Guzik}},\ }\bibfield
  {title} {\bibinfo {title} {Environment-assisted quantum walks in
  photosynthetic energy transfer},\ }\href@noop {} {\bibfield  {journal}
  {\bibinfo  {journal} {The Journal of chemical physics}\ }\textbf {\bibinfo
  {volume} {129}},\ \bibinfo {pages} {11B603} (\bibinfo {year}
  {2008})}\BibitemShut {NoStop}%
\bibitem [{\citenamefont {D'Acunto}(2021)}]{d2021protein}%
  \BibitemOpen
  \bibfield  {author} {\bibinfo {author} {\bibfnamefont {M.}~\bibnamefont
  {D'Acunto}},\ }\bibfield  {title} {\bibinfo {title} {Protein-{DNA} target
  search relies on quantum walk},\ }\href@noop {} {\bibfield  {journal}
  {\bibinfo  {journal} {Biosystems}\ }\textbf {\bibinfo {volume} {201}},\
  \bibinfo {pages} {104340} (\bibinfo {year} {2021})}\BibitemShut {NoStop}%
\bibitem [{\citenamefont {Di~Molfetta}\ \emph {et~al.}(2013)\citenamefont
  {Di~Molfetta}, \citenamefont {Brachet},\ and\ \citenamefont
  {Debbasch}}]{di2013quantum}%
  \BibitemOpen
  \bibfield  {author} {\bibinfo {author} {\bibfnamefont {G.}~\bibnamefont
  {Di~Molfetta}}, \bibinfo {author} {\bibfnamefont {M.}~\bibnamefont
  {Brachet}},\ and\ \bibinfo {author} {\bibfnamefont {F.}~\bibnamefont
  {Debbasch}},\ }\bibfield  {title} {\bibinfo {title} {Quantum walks as
  massless {D}irac fermions in curved space-time},\ }\href@noop {} {\bibfield
  {journal} {\bibinfo  {journal} {Physical Review A}\ }\textbf {\bibinfo
  {volume} {88}},\ \bibinfo {pages} {042301} (\bibinfo {year}
  {2013})}\BibitemShut {NoStop}%
\bibitem [{\citenamefont {Kitagawa}\ \emph {et~al.}(2010)\citenamefont
  {Kitagawa}, \citenamefont {Rudner}, \citenamefont {Berg},\ and\ \citenamefont
  {Demler}}]{kitagawa2010exploring}%
  \BibitemOpen
  \bibfield  {author} {\bibinfo {author} {\bibfnamefont {T.}~\bibnamefont
  {Kitagawa}}, \bibinfo {author} {\bibfnamefont {M.~S.}\ \bibnamefont
  {Rudner}}, \bibinfo {author} {\bibfnamefont {E.}~\bibnamefont {Berg}},\ and\
  \bibinfo {author} {\bibfnamefont {E.}~\bibnamefont {Demler}},\ }\bibfield
  {title} {\bibinfo {title} {Exploring topological phases with quantum walks},\
  }\href@noop {} {\bibfield  {journal} {\bibinfo  {journal} {Physical Review
  A}\ }\textbf {\bibinfo {volume} {82}},\ \bibinfo {pages} {033429} (\bibinfo
  {year} {2010})}\BibitemShut {NoStop}%
\bibitem [{\citenamefont {Obuse}\ and\ \citenamefont
  {Kawakami}(2011)}]{obuse2011topological}%
  \BibitemOpen
  \bibfield  {author} {\bibinfo {author} {\bibfnamefont {H.}~\bibnamefont
  {Obuse}}\ and\ \bibinfo {author} {\bibfnamefont {N.}~\bibnamefont
  {Kawakami}},\ }\bibfield  {title} {\bibinfo {title} {Topological phases and
  delocalization of quantum walks in random environments},\ }\href@noop {}
  {\bibfield  {journal} {\bibinfo  {journal} {Physical Review B}\ }\textbf
  {\bibinfo {volume} {84}},\ \bibinfo {pages} {195139} (\bibinfo {year}
  {2011})}\BibitemShut {NoStop}%
\bibitem [{\citenamefont {Mittal}\ \emph {et~al.}(2021)\citenamefont {Mittal},
  \citenamefont {Raj}, \citenamefont {Dey},\ and\ \citenamefont
  {Goyal}}]{mittal2021persistence}%
  \BibitemOpen
  \bibfield  {author} {\bibinfo {author} {\bibfnamefont {V.}~\bibnamefont
  {Mittal}}, \bibinfo {author} {\bibfnamefont {A.}~\bibnamefont {Raj}},
  \bibinfo {author} {\bibfnamefont {S.}~\bibnamefont {Dey}},\ and\ \bibinfo
  {author} {\bibfnamefont {S.~K.}\ \bibnamefont {Goyal}},\ }\bibfield  {title}
  {\bibinfo {title} {Persistence of topological phases in non-hermitian quantum
  walks},\ }\href@noop {} {\bibfield  {journal} {\bibinfo  {journal}
  {Scientific Reports}\ }\textbf {\bibinfo {volume} {11}},\ \bibinfo {pages}
  {10262} (\bibinfo {year} {2021})}\BibitemShut {NoStop}%
\bibitem [{\citenamefont {Mallick}\ \emph {et~al.}(2017)\citenamefont
  {Mallick}, \citenamefont {Mandal},\ and\ \citenamefont
  {Chandrashekar}}]{mallick2017neutrino}%
  \BibitemOpen
  \bibfield  {author} {\bibinfo {author} {\bibfnamefont {A.}~\bibnamefont
  {Mallick}}, \bibinfo {author} {\bibfnamefont {S.}~\bibnamefont {Mandal}},\
  and\ \bibinfo {author} {\bibfnamefont {C.}~\bibnamefont {Chandrashekar}},\
  }\bibfield  {title} {\bibinfo {title} {Neutrino oscillations in discrete-time
  quantum walk framework},\ }\href@noop {} {\bibfield  {journal} {\bibinfo
  {journal} {The European Physical Journal C}\ }\textbf {\bibinfo {volume}
  {77}},\ \bibinfo {pages} {1} (\bibinfo {year} {2017})}\BibitemShut {NoStop}%
\bibitem [{\citenamefont {Meyer}(1996)}]{meyer1996kt}%
  \BibitemOpen
  \bibfield  {author} {\bibinfo {author} {\bibfnamefont {D.~A.}\ \bibnamefont
  {Meyer}},\ }\bibfield  {title} {\bibinfo {title} {From quantum cellular
  automata to quantum lattice gases},\ }\href@noop {} {\bibfield  {journal}
  {\bibinfo  {journal} {Journal of Statistical Physics}\ }\textbf {\bibinfo
  {volume} {85}},\ \bibinfo {pages} {551} (\bibinfo {year} {1996})}\BibitemShut
  {NoStop}%
\bibitem [{\citenamefont {Farhi}\ and\ \citenamefont
  {Gutmann}(1998)}]{farhi1998quantum}%
  \BibitemOpen
  \bibfield  {author} {\bibinfo {author} {\bibfnamefont {E.}~\bibnamefont
  {Farhi}}\ and\ \bibinfo {author} {\bibfnamefont {S.}~\bibnamefont
  {Gutmann}},\ }\bibfield  {title} {\bibinfo {title} {Quantum computation and
  decision trees},\ }\href@noop {} {\bibfield  {journal} {\bibinfo  {journal}
  {Physical Review A}\ }\textbf {\bibinfo {volume} {58}},\ \bibinfo {pages}
  {915} (\bibinfo {year} {1998})}\BibitemShut {NoStop}%
\bibitem [{\citenamefont {Venegas-Andraca}(2012)}]{venegas2012quantum}%
  \BibitemOpen
  \bibfield  {author} {\bibinfo {author} {\bibfnamefont {S.~E.}\ \bibnamefont
  {Venegas-Andraca}},\ }\bibfield  {title} {\bibinfo {title} {Quantum walks: a
  comprehensive review},\ }\href@noop {} {\bibfield  {journal} {\bibinfo
  {journal} {Quantum Information Processing}\ }\textbf {\bibinfo {volume}
  {11}},\ \bibinfo {pages} {1015} (\bibinfo {year} {2012})}\BibitemShut
  {NoStop}%
\bibitem [{\citenamefont {Aharonov}\ \emph {et~al.}(2001)\citenamefont
  {Aharonov}, \citenamefont {Ambainis}, \citenamefont {Kempe},\ and\
  \citenamefont {Vazirani}}]{aharonov2001quantum}%
  \BibitemOpen
  \bibfield  {author} {\bibinfo {author} {\bibfnamefont {D.}~\bibnamefont
  {Aharonov}}, \bibinfo {author} {\bibfnamefont {A.}~\bibnamefont {Ambainis}},
  \bibinfo {author} {\bibfnamefont {J.}~\bibnamefont {Kempe}},\ and\ \bibinfo
  {author} {\bibfnamefont {U.}~\bibnamefont {Vazirani}},\ }\bibfield  {title}
  {\bibinfo {title} {Quantum walks on graphs},\ }\href@noop {} {\bibfield
  {journal} {\bibinfo  {journal} {Proceedings of the thirty-third annual ACM
  symposium on Theory of computing}\ ,\ \bibinfo {pages} {50}} (\bibinfo {year}
  {2001})}\BibitemShut {NoStop}%
\bibitem [{\citenamefont {Ambainis}\ \emph {et~al.}(2001)\citenamefont
  {Ambainis}, \citenamefont {Bach}, \citenamefont {Nayak}, \citenamefont
  {Vishwanath},\ and\ \citenamefont {Watrous}}]{ambainis2001one}%
  \BibitemOpen
  \bibfield  {author} {\bibinfo {author} {\bibfnamefont {A.}~\bibnamefont
  {Ambainis}}, \bibinfo {author} {\bibfnamefont {E.}~\bibnamefont {Bach}},
  \bibinfo {author} {\bibfnamefont {A.}~\bibnamefont {Nayak}}, \bibinfo
  {author} {\bibfnamefont {A.}~\bibnamefont {Vishwanath}},\ and\ \bibinfo
  {author} {\bibfnamefont {J.}~\bibnamefont {Watrous}},\ }\bibfield  {title}
  {\bibinfo {title} {One-dimensional quantum walks},\ }\href@noop {} {\bibfield
   {journal} {\bibinfo  {journal} {Proceedings of the thirty-third annual ACM
  symposium on Theory of computing}\ ,\ \bibinfo {pages} {37}} (\bibinfo {year}
  {2001})}\BibitemShut {NoStop}%
\bibitem [{\citenamefont {Kempe}(2003)}]{kempe2003quantum}%
  \BibitemOpen
  \bibfield  {author} {\bibinfo {author} {\bibfnamefont {J.}~\bibnamefont
  {Kempe}},\ }\bibfield  {title} {\bibinfo {title} {Quantum random walks: an
  introductory overview},\ }\href@noop {} {\bibfield  {journal} {\bibinfo
  {journal} {Contemporary Physics}\ }\textbf {\bibinfo {volume} {44}},\
  \bibinfo {pages} {307} (\bibinfo {year} {2003})}\BibitemShut {NoStop}%
\bibitem [{\citenamefont {Konno}(2002)}]{konno2002quantum}%
  \BibitemOpen
  \bibfield  {author} {\bibinfo {author} {\bibfnamefont {N.}~\bibnamefont
  {Konno}},\ }\bibfield  {title} {\bibinfo {title} {Quantum random walks in one
  dimension},\ }\href@noop {} {\bibfield  {journal} {\bibinfo  {journal}
  {Quantum Information Processing}\ }\textbf {\bibinfo {volume} {1}},\ \bibinfo
  {pages} {345} (\bibinfo {year} {2002})}\BibitemShut {NoStop}%
\bibitem [{\citenamefont {Venegas-Andraca}(2008)}]{venegas2008quantum}%
  \BibitemOpen
  \bibfield  {author} {\bibinfo {author} {\bibfnamefont {S.~E.}\ \bibnamefont
  {Venegas-Andraca}},\ }\bibfield  {title} {\bibinfo {title} {Quantum walks for
  computer scientists},\ }\href@noop {} {\bibfield  {journal} {\bibinfo
  {journal} {Synthesis Lectures on Quantum Computing}\ }\textbf {\bibinfo
  {volume} {1}},\ \bibinfo {pages} {1} (\bibinfo {year} {2008})}\BibitemShut
  {NoStop}%
\bibitem [{\citenamefont {Lavi{\v{c}}ka}\ \emph {et~al.}(2011)\citenamefont
  {Lavi{\v{c}}ka}, \citenamefont {Poto{\v{c}}ek}, \citenamefont {Kiss},
  \citenamefont {Lutz},\ and\ \citenamefont {Jex}}]{lavivcka2011quantum}%
  \BibitemOpen
  \bibfield  {author} {\bibinfo {author} {\bibfnamefont {H.}~\bibnamefont
  {Lavi{\v{c}}ka}}, \bibinfo {author} {\bibfnamefont {V.}~\bibnamefont
  {Poto{\v{c}}ek}}, \bibinfo {author} {\bibfnamefont {T.}~\bibnamefont {Kiss}},
  \bibinfo {author} {\bibfnamefont {E.}~\bibnamefont {Lutz}},\ and\ \bibinfo
  {author} {\bibfnamefont {I.}~\bibnamefont {Jex}},\ }\bibfield  {title}
  {\bibinfo {title} {Quantum walk with jumps},\ }\href@noop {} {\bibfield
  {journal} {\bibinfo  {journal} {The European Physical Journal D}\ }\textbf
  {\bibinfo {volume} {64}},\ \bibinfo {pages} {119} (\bibinfo {year}
  {2011})}\BibitemShut {NoStop}%
\bibitem [{\citenamefont {Pires}\ \emph {et~al.}(2019)\citenamefont {Pires},
  \citenamefont {Molfetta},\ and\ \citenamefont
  {Queir{\'o}s}}]{pires2019multiple}%
  \BibitemOpen
  \bibfield  {author} {\bibinfo {author} {\bibfnamefont {M.~A.}\ \bibnamefont
  {Pires}}, \bibinfo {author} {\bibfnamefont {G.~D.}\ \bibnamefont
  {Molfetta}},\ and\ \bibinfo {author} {\bibfnamefont {S.~M.~D.}\ \bibnamefont
  {Queir{\'o}s}},\ }\bibfield  {title} {\bibinfo {title} {Multiple transitions
  between normal and hyperballistic diffusion in quantum walks with
  time-dependent jumps},\ }\href@noop {} {\bibfield  {journal} {\bibinfo
  {journal} {Scientific Reports}\ }\textbf {\bibinfo {volume} {9}},\ \bibinfo
  {pages} {1} (\bibinfo {year} {2019})}\BibitemShut {NoStop}%
\bibitem [{\citenamefont {Mukhopadhyay}\ and\ \citenamefont
  {Sen}(2020)}]{mukhopadhyay2020persistent}%
  \BibitemOpen
  \bibfield  {author} {\bibinfo {author} {\bibfnamefont {S.}~\bibnamefont
  {Mukhopadhyay}}\ and\ \bibinfo {author} {\bibfnamefont {P.}~\bibnamefont
  {Sen}},\ }\bibfield  {title} {\bibinfo {title} {Persistent quantum walks:
  Dynamic phases and diverging timescales},\ }\href@noop {} {\bibfield
  {journal} {\bibinfo  {journal} {Physical Review Research}\ }\textbf {\bibinfo
  {volume} {2}},\ \bibinfo {pages} {023002} (\bibinfo {year}
  {2020})}\BibitemShut {NoStop}%
\bibitem [{\citenamefont {Pires}\ and\ \citenamefont
  {Queir{\'o}s}(2020)}]{pires2020quantum}%
  \BibitemOpen
  \bibfield  {author} {\bibinfo {author} {\bibfnamefont {M.~A.}\ \bibnamefont
  {Pires}}\ and\ \bibinfo {author} {\bibfnamefont {S.~D.}\ \bibnamefont
  {Queir{\'o}s}},\ }\bibfield  {title} {\bibinfo {title} {Quantum walks with
  sequential aperiodic jumps},\ }\href@noop {} {\bibfield  {journal} {\bibinfo
  {journal} {Physical Review E}\ }\textbf {\bibinfo {volume} {102}},\ \bibinfo
  {pages} {012104} (\bibinfo {year} {2020})}\BibitemShut {NoStop}%
\bibitem [{\citenamefont {Naves}\ \emph
  {et~al.}(2022{\natexlab{a}})\citenamefont {Naves}, \citenamefont {Pires},
  \citenamefont {Soares-Pinto},\ and\ \citenamefont
  {Queir{\'o}s}}]{naves2022enhancing}%
  \BibitemOpen
  \bibfield  {author} {\bibinfo {author} {\bibfnamefont {C.~B.}\ \bibnamefont
  {Naves}}, \bibinfo {author} {\bibfnamefont {M.~A.}\ \bibnamefont {Pires}},
  \bibinfo {author} {\bibfnamefont {D.~O.}\ \bibnamefont {Soares-Pinto}},\ and\
  \bibinfo {author} {\bibfnamefont {S.~M.~D.}\ \bibnamefont {Queir{\'o}s}},\
  }\bibfield  {title} {\bibinfo {title} {Enhancing entanglement with the
  generalized elephant quantum walk from localized and delocalized states},\
  }\href@noop {} {\bibfield  {journal} {\bibinfo  {journal} {Physical Review
  A}\ }\textbf {\bibinfo {volume} {106}},\ \bibinfo {pages} {042408} (\bibinfo
  {year} {2022}{\natexlab{a}})}\BibitemShut {NoStop}%
\bibitem [{\citenamefont {Salimi}\ and\ \citenamefont
  {Yosefjani}(2012)}]{salimi2012asymptotic}%
  \BibitemOpen
  \bibfield  {author} {\bibinfo {author} {\bibfnamefont {S.}~\bibnamefont
  {Salimi}}\ and\ \bibinfo {author} {\bibfnamefont {R.}~\bibnamefont
  {Yosefjani}},\ }\bibfield  {title} {\bibinfo {title} {Asymptotic entanglement
  in 1d quantum walks with a time-dependent coined},\ }\href@noop {} {\bibfield
   {journal} {\bibinfo  {journal} {International Journal of Modern Physics B}\
  }\textbf {\bibinfo {volume} {26}},\ \bibinfo {pages} {1250112} (\bibinfo
  {year} {2012})}\BibitemShut {NoStop}%
\bibitem [{\citenamefont {Vieira}\ \emph {et~al.}(2013)\citenamefont {Vieira},
  \citenamefont {Amorim},\ and\ \citenamefont
  {Rigolin}}]{vieira2013dynamically}%
  \BibitemOpen
  \bibfield  {author} {\bibinfo {author} {\bibfnamefont {R.}~\bibnamefont
  {Vieira}}, \bibinfo {author} {\bibfnamefont {E.~P.}\ \bibnamefont {Amorim}},\
  and\ \bibinfo {author} {\bibfnamefont {G.}~\bibnamefont {Rigolin}},\
  }\bibfield  {title} {\bibinfo {title} {Dynamically disordered quantum walk as
  a maximal entanglement generator},\ }\href@noop {} {\bibfield  {journal}
  {\bibinfo  {journal} {Physical Review Letters}\ }\textbf {\bibinfo {volume}
  {111}},\ \bibinfo {pages} {180503} (\bibinfo {year} {2013})}\BibitemShut
  {NoStop}%
\bibitem [{\citenamefont {Rohde}\ \emph {et~al.}(2013)\citenamefont {Rohde},
  \citenamefont {Brennen},\ and\ \citenamefont {Gilchrist}}]{rohde2013quantum}%
  \BibitemOpen
  \bibfield  {author} {\bibinfo {author} {\bibfnamefont {P.~P.}\ \bibnamefont
  {Rohde}}, \bibinfo {author} {\bibfnamefont {G.~K.}\ \bibnamefont {Brennen}},\
  and\ \bibinfo {author} {\bibfnamefont {A.}~\bibnamefont {Gilchrist}},\
  }\bibfield  {title} {\bibinfo {title} {Quantum walks with memory provided by
  recycled coins and a memory of the coin-flip history},\ }\href@noop {}
  {\bibfield  {journal} {\bibinfo  {journal} {Physical Review A}\ }\textbf
  {\bibinfo {volume} {87}},\ \bibinfo {pages} {052302} (\bibinfo {year}
  {2013})}\BibitemShut {NoStop}%
\bibitem [{\citenamefont {Vieira}\ \emph {et~al.}(2014)\citenamefont {Vieira},
  \citenamefont {Amorim},\ and\ \citenamefont
  {Rigolin}}]{vieira2014entangling}%
  \BibitemOpen
  \bibfield  {author} {\bibinfo {author} {\bibfnamefont {R.}~\bibnamefont
  {Vieira}}, \bibinfo {author} {\bibfnamefont {E.~P.}\ \bibnamefont {Amorim}},\
  and\ \bibinfo {author} {\bibfnamefont {G.}~\bibnamefont {Rigolin}},\
  }\bibfield  {title} {\bibinfo {title} {Entangling power of disordered quantum
  walks},\ }\href@noop {} {\bibfield  {journal} {\bibinfo  {journal} {Physical
  Review A}\ }\textbf {\bibinfo {volume} {89}},\ \bibinfo {pages} {042307}
  (\bibinfo {year} {2014})}\BibitemShut {NoStop}%
\bibitem [{\citenamefont {Di~Molfetta}\ and\ \citenamefont
  {Debbasch}(2016)}]{di2016discrete}%
  \BibitemOpen
  \bibfield  {author} {\bibinfo {author} {\bibfnamefont {G.}~\bibnamefont
  {Di~Molfetta}}\ and\ \bibinfo {author} {\bibfnamefont {F.}~\bibnamefont
  {Debbasch}},\ }\bibfield  {title} {\bibinfo {title} {Discrete-time quantum
  walks in random artificial gauge fields},\ }\href@noop {} {\bibfield
  {journal} {\bibinfo  {journal} {Quantum Studies: Mathematics and
  Foundations}\ }\textbf {\bibinfo {volume} {3}},\ \bibinfo {pages} {293}
  (\bibinfo {year} {2016})}\BibitemShut {NoStop}%
\bibitem [{\citenamefont {Montero}(2016)}]{montero2016classical}%
  \BibitemOpen
  \bibfield  {author} {\bibinfo {author} {\bibfnamefont {M.}~\bibnamefont
  {Montero}},\ }\bibfield  {title} {\bibinfo {title} {Classical-like behavior
  in quantum walks with inhomogeneous, time-dependent coin operators},\
  }\href@noop {} {\bibfield  {journal} {\bibinfo  {journal} {Physical Review
  A}\ }\textbf {\bibinfo {volume} {93}},\ \bibinfo {pages} {062316} (\bibinfo
  {year} {2016})}\BibitemShut {NoStop}%
\bibitem [{\citenamefont {Wang}\ \emph {et~al.}(2018)\citenamefont {Wang},
  \citenamefont {Xu}, \citenamefont {Pan}, \citenamefont {Sun}, \citenamefont
  {Xu}, \citenamefont {Chen}, \citenamefont {Han}, \citenamefont {Li},\ and\
  \citenamefont {Guo}}]{wang2018dynamic}%
  \BibitemOpen
  \bibfield  {author} {\bibinfo {author} {\bibfnamefont {Q.-Q.}\ \bibnamefont
  {Wang}}, \bibinfo {author} {\bibfnamefont {X.-Y.}\ \bibnamefont {Xu}},
  \bibinfo {author} {\bibfnamefont {W.-W.}\ \bibnamefont {Pan}}, \bibinfo
  {author} {\bibfnamefont {K.}~\bibnamefont {Sun}}, \bibinfo {author}
  {\bibfnamefont {J.-S.}\ \bibnamefont {Xu}}, \bibinfo {author} {\bibfnamefont
  {G.}~\bibnamefont {Chen}}, \bibinfo {author} {\bibfnamefont {Y.-J.}\
  \bibnamefont {Han}}, \bibinfo {author} {\bibfnamefont {C.-F.}\ \bibnamefont
  {Li}},\ and\ \bibinfo {author} {\bibfnamefont {G.-C.}\ \bibnamefont {Guo}},\
  }\bibfield  {title} {\bibinfo {title} {Dynamic-disorder-induced enhancement
  of entanglement in photonic quantum walks},\ }\href@noop {} {\bibfield
  {journal} {\bibinfo  {journal} {Optica}\ }\textbf {\bibinfo {volume} {5}},\
  \bibinfo {pages} {1136} (\bibinfo {year} {2018})}\BibitemShut {NoStop}%
\bibitem [{\citenamefont {Orthey}\ and\ \citenamefont
  {Amorim}(2019)}]{orthey2019weak}%
  \BibitemOpen
  \bibfield  {author} {\bibinfo {author} {\bibfnamefont {A.~C.}\ \bibnamefont
  {Orthey}}\ and\ \bibinfo {author} {\bibfnamefont {E.~P.}\ \bibnamefont
  {Amorim}},\ }\bibfield  {title} {\bibinfo {title} {Weak disorder enhancing
  the production of entanglement in quantum walks},\ }\href@noop {} {\bibfield
  {journal} {\bibinfo  {journal} {Brazilian Journal of Physics}\ }\textbf
  {\bibinfo {volume} {49}},\ \bibinfo {pages} {595} (\bibinfo {year}
  {2019})}\BibitemShut {NoStop}%
\bibitem [{\citenamefont {Singh}\ \emph {et~al.}(2019)\citenamefont {Singh},
  \citenamefont {Balu}, \citenamefont {Laflamme},\ and\ \citenamefont
  {Chandrashekar}}]{singh2019accelerated}%
  \BibitemOpen
  \bibfield  {author} {\bibinfo {author} {\bibfnamefont {S.}~\bibnamefont
  {Singh}}, \bibinfo {author} {\bibfnamefont {R.}~\bibnamefont {Balu}},
  \bibinfo {author} {\bibfnamefont {R.}~\bibnamefont {Laflamme}},\ and\
  \bibinfo {author} {\bibfnamefont {C.}~\bibnamefont {Chandrashekar}},\
  }\bibfield  {title} {\bibinfo {title} {Accelerated quantum walk, two-particle
  entanglement generation and localization},\ }\href@noop {} {\bibfield
  {journal} {\bibinfo  {journal} {Journal of Physics Communications}\ }\textbf
  {\bibinfo {volume} {3}},\ \bibinfo {pages} {055008} (\bibinfo {year}
  {2019})}\BibitemShut {NoStop}%
\bibitem [{\citenamefont {Buarque}\ and\ \citenamefont
  {Dias}(2019)}]{buarque2019aperiodic}%
  \BibitemOpen
  \bibfield  {author} {\bibinfo {author} {\bibfnamefont {A.}~\bibnamefont
  {Buarque}}\ and\ \bibinfo {author} {\bibfnamefont {W.~d.~S.}\ \bibnamefont
  {Dias}},\ }\bibfield  {title} {\bibinfo {title} {Aperiodic
  space-inhomogeneous quantum walks: Localization properties, energy spectra,
  and enhancement of entanglement},\ }\href@noop {} {\bibfield  {journal}
  {\bibinfo  {journal} {Physical Review E}\ }\textbf {\bibinfo {volume}
  {100}},\ \bibinfo {pages} {032106} (\bibinfo {year} {2019})}\BibitemShut
  {NoStop}%
\bibitem [{\citenamefont {Pires}\ and\ \citenamefont
  {Duarte~Queir{\'o}s}(2021)}]{pires2021negative}%
  \BibitemOpen
  \bibfield  {author} {\bibinfo {author} {\bibfnamefont {M.~A.}\ \bibnamefont
  {Pires}}\ and\ \bibinfo {author} {\bibfnamefont {S.~M.}\ \bibnamefont
  {Duarte~Queir{\'o}s}},\ }\bibfield  {title} {\bibinfo {title} {Negative
  correlations can play a positive role in disordered quantum walks},\
  }\href@noop {} {\bibfield  {journal} {\bibinfo  {journal} {Scientific
  Reports}\ }\textbf {\bibinfo {volume} {11}},\ \bibinfo {pages} {4527}
  (\bibinfo {year} {2021})}\BibitemShut {NoStop}%
\bibitem [{\citenamefont {Naves}\ \emph {et~al.}(2023)\citenamefont {Naves},
  \citenamefont {Pires}, \citenamefont {Soares-Pinto},\ and\ \citenamefont
  {Queir{\'o}s}}]{naves2023quantum}%
  \BibitemOpen
  \bibfield  {author} {\bibinfo {author} {\bibfnamefont {C.~B.}\ \bibnamefont
  {Naves}}, \bibinfo {author} {\bibfnamefont {M.~A.}\ \bibnamefont {Pires}},
  \bibinfo {author} {\bibfnamefont {D.~O.}\ \bibnamefont {Soares-Pinto}},\ and\
  \bibinfo {author} {\bibfnamefont {S.~M.~D.}\ \bibnamefont {Queir{\'o}s}},\
  }\bibfield  {title} {\bibinfo {title} {Quantum walks in two dimensions:
  controlling directional spreading with entangling coins and tunable
  disordered step operator},\ }\href@noop {} {\bibfield  {journal} {\bibinfo
  {journal} {Journal of Physics A: Mathematical and Theoretical}\ }\textbf
  {\bibinfo {volume} {56}},\ \bibinfo {pages} {125301} (\bibinfo {year}
  {2023})}\BibitemShut {NoStop}%
\bibitem [{\citenamefont {Linden}\ and\ \citenamefont
  {Sharam}(2009)}]{linden2009inhomogeneous}%
  \BibitemOpen
  \bibfield  {author} {\bibinfo {author} {\bibfnamefont {N.}~\bibnamefont
  {Linden}}\ and\ \bibinfo {author} {\bibfnamefont {J.}~\bibnamefont
  {Sharam}},\ }\bibfield  {title} {\bibinfo {title} {Inhomogeneous quantum
  walks},\ }\href@noop {} {\bibfield  {journal} {\bibinfo  {journal} {Physical
  Review A}\ }\textbf {\bibinfo {volume} {80}},\ \bibinfo {pages} {052327}
  (\bibinfo {year} {2009})}\BibitemShut {NoStop}%
\bibitem [{\citenamefont {Brun}\ \emph {et~al.}(2003)\citenamefont {Brun},
  \citenamefont {Carteret},\ and\ \citenamefont {Ambainis}}]{brun2003quantum}%
  \BibitemOpen
  \bibfield  {author} {\bibinfo {author} {\bibfnamefont {T.~A.}\ \bibnamefont
  {Brun}}, \bibinfo {author} {\bibfnamefont {H.~A.}\ \bibnamefont {Carteret}},\
  and\ \bibinfo {author} {\bibfnamefont {A.}~\bibnamefont {Ambainis}},\
  }\bibfield  {title} {\bibinfo {title} {Quantum walks driven by many coins},\
  }\href@noop {} {\bibfield  {journal} {\bibinfo  {journal} {Physical Review
  A}\ }\textbf {\bibinfo {volume} {67}},\ \bibinfo {pages} {052317} (\bibinfo
  {year} {2003})}\BibitemShut {NoStop}%
\bibitem [{\citenamefont {Oliveira}\ \emph {et~al.}(2006)\citenamefont
  {Oliveira}, \citenamefont {Portugal},\ and\ \citenamefont
  {Donangelo}}]{oliveira2006decoherence}%
  \BibitemOpen
  \bibfield  {author} {\bibinfo {author} {\bibfnamefont {A.}~\bibnamefont
  {Oliveira}}, \bibinfo {author} {\bibfnamefont {R.}~\bibnamefont {Portugal}},\
  and\ \bibinfo {author} {\bibfnamefont {R.}~\bibnamefont {Donangelo}},\
  }\bibfield  {title} {\bibinfo {title} {Decoherence in two-dimensional quantum
  walks},\ }\href@noop {} {\bibfield  {journal} {\bibinfo  {journal} {Physical
  Review A}\ }\textbf {\bibinfo {volume} {74}},\ \bibinfo {pages} {012312}
  (\bibinfo {year} {2006})}\BibitemShut {NoStop}%
\bibitem [{\citenamefont {G{\"o}n{\"u}lol}\ \emph {et~al.}(2009)\citenamefont
  {G{\"o}n{\"u}lol}, \citenamefont {Aydiner},\ and\ \citenamefont
  {M{\"u}stecapl{\i}o{\u{g}}lu}}]{gonulol2009decoherence}%
  \BibitemOpen
  \bibfield  {author} {\bibinfo {author} {\bibfnamefont {M.}~\bibnamefont
  {G{\"o}n{\"u}lol}}, \bibinfo {author} {\bibfnamefont {E.}~\bibnamefont
  {Aydiner}},\ and\ \bibinfo {author} {\bibfnamefont {{\"O}.~E.}\ \bibnamefont
  {M{\"u}stecapl{\i}o{\u{g}}lu}},\ }\bibfield  {title} {\bibinfo {title}
  {Decoherence in two-dimensional quantum random walks with traps},\
  }\href@noop {} {\bibfield  {journal} {\bibinfo  {journal} {Physical Review
  A}\ }\textbf {\bibinfo {volume} {80}},\ \bibinfo {pages} {022336} (\bibinfo
  {year} {2009})}\BibitemShut {NoStop}%
\bibitem [{\citenamefont {Chandrashekar}\ and\ \citenamefont
  {Busch}(2013)}]{chandrashekar2013decoherence}%
  \BibitemOpen
  \bibfield  {author} {\bibinfo {author} {\bibfnamefont {C.}~\bibnamefont
  {Chandrashekar}}\ and\ \bibinfo {author} {\bibfnamefont {T.}~\bibnamefont
  {Busch}},\ }\bibfield  {title} {\bibinfo {title} {Decoherence in
  two-dimensional quantum walks using four-and two-state particles},\
  }\href@noop {} {\bibfield  {journal} {\bibinfo  {journal} {Journal of Physics
  A: Mathematical and Theoretical}\ }\textbf {\bibinfo {volume} {46}},\
  \bibinfo {pages} {105306} (\bibinfo {year} {2013})}\BibitemShut {NoStop}%
\bibitem [{\citenamefont {Annabestani}\ \emph {et~al.}(2010)\citenamefont
  {Annabestani}, \citenamefont {Akhtarshenas},\ and\ \citenamefont
  {Abolhassani}}]{annabestani2010decoherence}%
  \BibitemOpen
  \bibfield  {author} {\bibinfo {author} {\bibfnamefont {M.}~\bibnamefont
  {Annabestani}}, \bibinfo {author} {\bibfnamefont {S.~J.}\ \bibnamefont
  {Akhtarshenas}},\ and\ \bibinfo {author} {\bibfnamefont {M.~R.}\ \bibnamefont
  {Abolhassani}},\ }\bibfield  {title} {\bibinfo {title} {Decoherence in a
  one-dimensional quantum walk},\ }\href@noop {} {\bibfield  {journal}
  {\bibinfo  {journal} {Physical Review A}\ }\textbf {\bibinfo {volume} {81}},\
  \bibinfo {pages} {032321} (\bibinfo {year} {2010})}\BibitemShut {NoStop}%
\bibitem [{\citenamefont {Ghosal}\ and\ \citenamefont
  {Deb}(2018)}]{PhysRevA.98.032104}%
  \BibitemOpen
  \bibfield  {author} {\bibinfo {author} {\bibfnamefont {A.}~\bibnamefont
  {Ghosal}}\ and\ \bibinfo {author} {\bibfnamefont {P.}~\bibnamefont {Deb}},\
  }\bibfield  {title} {\bibinfo {title} {Quantum walks over a square lattice},\
  }\href {https://doi.org/10.1103/PhysRevA.98.032104} {\bibfield  {journal}
  {\bibinfo  {journal} {Phys. Rev. A}\ }\textbf {\bibinfo {volume} {98}},\
  \bibinfo {pages} {032104} (\bibinfo {year} {2018})}\BibitemShut {NoStop}%
\bibitem [{\citenamefont {Yang}\ \emph {et~al.}(2021)\citenamefont {Yang},
  \citenamefont {Wang}, \citenamefont {Li}, \citenamefont {Li}, \citenamefont
  {Zhou},\ and\ \citenamefont {Shi}}]{yang2021decoherence}%
  \BibitemOpen
  \bibfield  {author} {\bibinfo {author} {\bibfnamefont {Y.-G.}\ \bibnamefont
  {Yang}}, \bibinfo {author} {\bibfnamefont {X.-X.}\ \bibnamefont {Wang}},
  \bibinfo {author} {\bibfnamefont {J.}~\bibnamefont {Li}}, \bibinfo {author}
  {\bibfnamefont {D.}~\bibnamefont {Li}}, \bibinfo {author} {\bibfnamefont
  {Y.-H.}\ \bibnamefont {Zhou}},\ and\ \bibinfo {author} {\bibfnamefont
  {W.-M.}\ \bibnamefont {Shi}},\ }\bibfield  {title} {\bibinfo {title}
  {Decoherence in two-dimensional quantum walks with two-and four-state
  coins},\ }\href@noop {} {\bibfield  {journal} {\bibinfo  {journal} {Modern
  Physics Letters A}\ }\textbf {\bibinfo {volume} {36}},\ \bibinfo {pages}
  {2150210} (\bibinfo {year} {2021})}\BibitemShut {NoStop}%
\bibitem [{\citenamefont {Tude}\ and\ \citenamefont
  {de~Oliveira}(2022)}]{tude2022decoherence}%
  \BibitemOpen
  \bibfield  {author} {\bibinfo {author} {\bibfnamefont {L.~T.}\ \bibnamefont
  {Tude}}\ and\ \bibinfo {author} {\bibfnamefont {M.~C.}\ \bibnamefont
  {de~Oliveira}},\ }\bibfield  {title} {\bibinfo {title} {Decoherence in the
  three-state quantum walk},\ }\href@noop {} {\bibfield  {journal} {\bibinfo
  {journal} {Physica A: Statistical Mechanics and its Applications}\ }\textbf
  {\bibinfo {volume} {605}},\ \bibinfo {pages} {128012} (\bibinfo {year}
  {2022})}\BibitemShut {NoStop}%
\bibitem [{\citenamefont {Xue}\ \emph {et~al.}(2015)\citenamefont {Xue},
  \citenamefont {Zhang}, \citenamefont {Bian}, \citenamefont {Zhan},
  \citenamefont {Qin},\ and\ \citenamefont {Sanders}}]{xue2015localized}%
  \BibitemOpen
  \bibfield  {author} {\bibinfo {author} {\bibfnamefont {P.}~\bibnamefont
  {Xue}}, \bibinfo {author} {\bibfnamefont {R.}~\bibnamefont {Zhang}}, \bibinfo
  {author} {\bibfnamefont {Z.}~\bibnamefont {Bian}}, \bibinfo {author}
  {\bibfnamefont {X.}~\bibnamefont {Zhan}}, \bibinfo {author} {\bibfnamefont
  {H.}~\bibnamefont {Qin}},\ and\ \bibinfo {author} {\bibfnamefont {B.~C.}\
  \bibnamefont {Sanders}},\ }\bibfield  {title} {\bibinfo {title} {Localized
  state in a two-dimensional quantum walk on a disordered lattice},\
  }\href@noop {} {\bibfield  {journal} {\bibinfo  {journal} {Physical Review
  A}\ }\textbf {\bibinfo {volume} {92}},\ \bibinfo {pages} {042316} (\bibinfo
  {year} {2015})}\BibitemShut {NoStop}%
\bibitem [{\citenamefont {Di~Molfetta}\ \emph {et~al.}(2018)\citenamefont
  {Di~Molfetta}, \citenamefont {Soares-Pinto},\ and\ \citenamefont
  {Queir{\'o}s}}]{di2018elephant}%
  \BibitemOpen
  \bibfield  {author} {\bibinfo {author} {\bibfnamefont {G.}~\bibnamefont
  {Di~Molfetta}}, \bibinfo {author} {\bibfnamefont {D.~O.}\ \bibnamefont
  {Soares-Pinto}},\ and\ \bibinfo {author} {\bibfnamefont {S.~M.~D.}\
  \bibnamefont {Queir{\'o}s}},\ }\bibfield  {title} {\bibinfo {title} {Elephant
  quantum walk},\ }\href@noop {} {\bibfield  {journal} {\bibinfo  {journal}
  {Physical Review A}\ }\textbf {\bibinfo {volume} {97}},\ \bibinfo {pages}
  {062112} (\bibinfo {year} {2018})}\BibitemShut {NoStop}%
\bibitem [{\citenamefont {Naves}\ \emph
  {et~al.}(2022{\natexlab{b}})\citenamefont {Naves}, \citenamefont {Pires},
  \citenamefont {Soares-Pinto},\ and\ \citenamefont
  {Duarte~Queir{\'o}s}}]{naves2022quantum}%
  \BibitemOpen
  \bibfield  {author} {\bibinfo {author} {\bibfnamefont {C.}~\bibnamefont
  {Naves}}, \bibinfo {author} {\bibfnamefont {M.~A.}\ \bibnamefont {Pires}},
  \bibinfo {author} {\bibfnamefont {D.~O.}\ \bibnamefont {Soares-Pinto}},\ and\
  \bibinfo {author} {\bibfnamefont {S.~M.}\ \bibnamefont
  {Duarte~Queir{\'o}s}},\ }\bibfield  {title} {\bibinfo {title} {Quantum walks
  in two dimensions: controlling directional spreading with entangling coins
  and tunable disordered step operator},\ }\href@noop {} {\bibfield  {journal}
  {\bibinfo  {journal} {Journal of Physics A: Mathematical and Theoretical}\ }
  (\bibinfo {year} {2022}{\natexlab{b}})}\BibitemShut {NoStop}%
\bibitem [{\citenamefont {Chatterjee}\ and\ \citenamefont
  {Loring}(1994)}]{chatterjee1994effective}%
  \BibitemOpen
  \bibfield  {author} {\bibinfo {author} {\bibfnamefont {A.~P.}\ \bibnamefont
  {Chatterjee}}\ and\ \bibinfo {author} {\bibfnamefont {R.~F.}\ \bibnamefont
  {Loring}},\ }\bibfield  {title} {\bibinfo {title} {Effective medium
  approximation for random walks with non-markovian dynamical disorder},\
  }\href@noop {} {\bibfield  {journal} {\bibinfo  {journal} {Physical Review
  E}\ }\textbf {\bibinfo {volume} {50}},\ \bibinfo {pages} {2439} (\bibinfo
  {year} {1994})}\BibitemShut {NoStop}%
\bibitem [{\citenamefont {Yin}\ \emph {et~al.}(2008)\citenamefont {Yin},
  \citenamefont {Katsanos},\ and\ \citenamefont {Evangelou}}]{yin2008quantum}%
  \BibitemOpen
  \bibfield  {author} {\bibinfo {author} {\bibfnamefont {Y.}~\bibnamefont
  {Yin}}, \bibinfo {author} {\bibfnamefont {D.}~\bibnamefont {Katsanos}},\ and\
  \bibinfo {author} {\bibfnamefont {S.}~\bibnamefont {Evangelou}},\ }\bibfield
  {title} {\bibinfo {title} {Quantum walks on a random environment},\
  }\href@noop {} {\bibfield  {journal} {\bibinfo  {journal} {Physical Review
  A}\ }\textbf {\bibinfo {volume} {77}},\ \bibinfo {pages} {022302} (\bibinfo
  {year} {2008})}\BibitemShut {NoStop}%
\bibitem [{\citenamefont {Nosrati}\ \emph {et~al.}(2021)\citenamefont
  {Nosrati}, \citenamefont {Laneve}, \citenamefont {Shadfar}, \citenamefont
  {Geraldi}, \citenamefont {Mahdavipour}, \citenamefont {Pegoraro},
  \citenamefont {Mataloni},\ and\ \citenamefont {Franco}}]{nosrati2021readout}%
  \BibitemOpen
  \bibfield  {author} {\bibinfo {author} {\bibfnamefont {F.}~\bibnamefont
  {Nosrati}}, \bibinfo {author} {\bibfnamefont {A.}~\bibnamefont {Laneve}},
  \bibinfo {author} {\bibfnamefont {M.~K.}\ \bibnamefont {Shadfar}}, \bibinfo
  {author} {\bibfnamefont {A.}~\bibnamefont {Geraldi}}, \bibinfo {author}
  {\bibfnamefont {K.}~\bibnamefont {Mahdavipour}}, \bibinfo {author}
  {\bibfnamefont {F.}~\bibnamefont {Pegoraro}}, \bibinfo {author}
  {\bibfnamefont {P.}~\bibnamefont {Mataloni}},\ and\ \bibinfo {author}
  {\bibfnamefont {R.~L.}\ \bibnamefont {Franco}},\ }\bibfield  {title}
  {\bibinfo {title} {Readout of quantum information spreading using a
  disordered quantum walk},\ }\href@noop {} {\bibfield  {journal} {\bibinfo
  {journal} {JOSA B}\ }\textbf {\bibinfo {volume} {38}},\ \bibinfo {pages}
  {2570} (\bibinfo {year} {2021})}\BibitemShut {NoStop}%
\bibitem [{\citenamefont {P{\'o}lya}(1921)}]{polya1921aufgabe}%
  \BibitemOpen
  \bibfield  {author} {\bibinfo {author} {\bibfnamefont {G.}~\bibnamefont
  {P{\'o}lya}},\ }\bibfield  {title} {\bibinfo {title} {{\"U}ber eine aufgabe
  der wahrscheinlichkeitsrechnung betreffend die irrfahrt im stra{\ss}ennetz},\
  }\href@noop {} {\bibfield  {journal} {\bibinfo  {journal} {Mathematische
  Annalen}\ }\textbf {\bibinfo {volume} {84}},\ \bibinfo {pages} {149}
  (\bibinfo {year} {1921})}\BibitemShut {NoStop}%
\bibitem [{\citenamefont {R{\'e}v{\'e}sz}(2013)}]{revesz2013random}%
  \BibitemOpen
  \bibfield  {author} {\bibinfo {author} {\bibfnamefont {P.}~\bibnamefont
  {R{\'e}v{\'e}sz}},\ }\bibfield  {title} {\bibinfo {title} {Random walk in
  random and non-random environments},\ }\href@noop {} {\bibfield  {journal}
  {\bibinfo  {journal} {World Scientific}\ } (\bibinfo {year}
  {2013})}\BibitemShut {NoStop}%
\bibitem [{\citenamefont {Nayak}\ and\ \citenamefont
  {Vishwanath}(2000)}]{nayak2000quantum}%
  \BibitemOpen
  \bibfield  {author} {\bibinfo {author} {\bibfnamefont {A.}~\bibnamefont
  {Nayak}}\ and\ \bibinfo {author} {\bibfnamefont {A.}~\bibnamefont
  {Vishwanath}},\ }\bibfield  {title} {\bibinfo {title} {Quantum walk on the
  line},\ }\href@noop {} {\bibfield  {journal} {\bibinfo  {journal} {arXiv
  preprint quant-ph/0010117}\ } (\bibinfo {year} {2000})}\BibitemShut {NoStop}%
\bibitem [{\citenamefont {Portugal}(2013)}]{portugal2013quantum}%
  \BibitemOpen
  \bibfield  {author} {\bibinfo {author} {\bibfnamefont {R.}~\bibnamefont
  {Portugal}},\ }\bibfield  {title} {\bibinfo {title} {Quantum walks and search
  algorithms, quantum science and technology, vol. 19},\ }\href@noop {}
  {\bibfield  {journal} {\bibinfo  {journal} {Springer}\ } (\bibinfo {year}
  {2013})}\BibitemShut {NoStop}%
\bibitem [{\citenamefont {Inui}\ \emph {et~al.}(2004)\citenamefont {Inui},
  \citenamefont {Konishi},\ and\ \citenamefont {Konno}}]{inui2004localization}%
  \BibitemOpen
  \bibfield  {author} {\bibinfo {author} {\bibfnamefont {N.}~\bibnamefont
  {Inui}}, \bibinfo {author} {\bibfnamefont {Y.}~\bibnamefont {Konishi}},\ and\
  \bibinfo {author} {\bibfnamefont {N.}~\bibnamefont {Konno}},\ }\bibfield
  {title} {\bibinfo {title} {Localization of two-dimensional quantum walks},\
  }\href@noop {} {\bibfield  {journal} {\bibinfo  {journal} {Physical Review
  A}\ }\textbf {\bibinfo {volume} {69}},\ \bibinfo {pages} {052323} (\bibinfo
  {year} {2004})}\BibitemShut {NoStop}%
\bibitem [{\citenamefont {{\v{S}}tefa{\v{n}}{\'a}k}\ \emph
  {et~al.}(2008)\citenamefont {{\v{S}}tefa{\v{n}}{\'a}k}, \citenamefont
  {Kiss},\ and\ \citenamefont {Jex}}]{vstefavnak2008recurrence}%
  \BibitemOpen
  \bibfield  {author} {\bibinfo {author} {\bibfnamefont {M.}~\bibnamefont
  {{\v{S}}tefa{\v{n}}{\'a}k}}, \bibinfo {author} {\bibfnamefont
  {T.}~\bibnamefont {Kiss}},\ and\ \bibinfo {author} {\bibfnamefont
  {I.}~\bibnamefont {Jex}},\ }\bibfield  {title} {\bibinfo {title} {Recurrence
  properties of unbiased coined quantum walks on infinite d-dimensional
  lattices},\ }\href@noop {} {\bibfield  {journal} {\bibinfo  {journal}
  {Physical Review A}\ }\textbf {\bibinfo {volume} {78}},\ \bibinfo {pages}
  {032306} (\bibinfo {year} {2008})}\BibitemShut {NoStop}%
\bibitem [{\citenamefont {Grover}(1996)}]{grover1996fast}%
  \BibitemOpen
  \bibfield  {author} {\bibinfo {author} {\bibfnamefont {L.~K.}\ \bibnamefont
  {Grover}},\ }\bibfield  {title} {\bibinfo {title} {A fast quantum mechanical
  algorithm for database search},\ }\href@noop {} {\bibfield  {journal}
  {\bibinfo  {journal} {Proceedings of the twenty-eighth annual ACM symposium
  on Theory of computing}\ ,\ \bibinfo {pages} {212}} (\bibinfo {year}
  {1996})}\BibitemShut {NoStop}%
\bibitem [{\citenamefont {Mackay}\ \emph {et~al.}(2002)\citenamefont {Mackay},
  \citenamefont {Bartlett}, \citenamefont {Stephenson},\ and\ \citenamefont
  {Sanders}}]{mackay2002quantum}%
  \BibitemOpen
  \bibfield  {author} {\bibinfo {author} {\bibfnamefont {T.~D.}\ \bibnamefont
  {Mackay}}, \bibinfo {author} {\bibfnamefont {S.~D.}\ \bibnamefont
  {Bartlett}}, \bibinfo {author} {\bibfnamefont {L.~T.}\ \bibnamefont
  {Stephenson}},\ and\ \bibinfo {author} {\bibfnamefont {B.~C.}\ \bibnamefont
  {Sanders}},\ }\bibfield  {title} {\bibinfo {title} {Quantum walks in higher
  dimensions},\ }\href@noop {} {\bibfield  {journal} {\bibinfo  {journal}
  {Journal of Physics A: Mathematical and General}\ }\textbf {\bibinfo {volume}
  {35}},\ \bibinfo {pages} {2745} (\bibinfo {year} {2002})}\BibitemShut
  {NoStop}%
\bibitem [{\citenamefont {Di~Franco}\ \emph {et~al.}(2011)\citenamefont
  {Di~Franco}, \citenamefont {Mc~Gettrick},\ and\ \citenamefont
  {Busch}}]{di2011mimicking}%
  \BibitemOpen
  \bibfield  {author} {\bibinfo {author} {\bibfnamefont {C.}~\bibnamefont
  {Di~Franco}}, \bibinfo {author} {\bibfnamefont {M.}~\bibnamefont
  {Mc~Gettrick}},\ and\ \bibinfo {author} {\bibfnamefont {T.}~\bibnamefont
  {Busch}},\ }\bibfield  {title} {\bibinfo {title} {Mimicking the probability
  distribution of a two-dimensional grover walk with a single-qubit coin},\
  }\href@noop {} {\bibfield  {journal} {\bibinfo  {journal} {Physical Review
  Letters}\ }\textbf {\bibinfo {volume} {106}},\ \bibinfo {pages} {080502}
  (\bibinfo {year} {2011})}\BibitemShut {NoStop}%
\bibitem [{\citenamefont {Grimmett}\ \emph {et~al.}(2004)\citenamefont
  {Grimmett}, \citenamefont {Janson},\ and\ \citenamefont
  {Scudo}}]{grimmett2004weak}%
  \BibitemOpen
  \bibfield  {author} {\bibinfo {author} {\bibfnamefont {G.}~\bibnamefont
  {Grimmett}}, \bibinfo {author} {\bibfnamefont {S.}~\bibnamefont {Janson}},\
  and\ \bibinfo {author} {\bibfnamefont {P.~F.}\ \bibnamefont {Scudo}},\
  }\bibfield  {title} {\bibinfo {title} {Weak limits for quantum random
  walks},\ }\href@noop {} {\bibfield  {journal} {\bibinfo  {journal} {Physical
  Review E}\ }\textbf {\bibinfo {volume} {69}},\ \bibinfo {pages} {026119}
  (\bibinfo {year} {2004})}\BibitemShut {NoStop}%
\bibitem [{\citenamefont {Konno}(2005)}]{konno2005new}%
  \BibitemOpen
  \bibfield  {author} {\bibinfo {author} {\bibfnamefont {N.}~\bibnamefont
  {Konno}},\ }\bibfield  {title} {\bibinfo {title} {A new type of limit
  theorems for the one-dimensional quantum random walk},\ }\href@noop {}
  {\bibfield  {journal} {\bibinfo  {journal} {Journal of the Mathematical
  Society of Japan}\ }\textbf {\bibinfo {volume} {57}},\ \bibinfo {pages}
  {1179} (\bibinfo {year} {2005})}\BibitemShut {NoStop}%
\bibitem [{\citenamefont {Carleo}\ \emph {et~al.}(2012)\citenamefont {Carleo},
  \citenamefont {Becca}, \citenamefont {Schir{\'o}},\ and\ \citenamefont
  {Fabrizio}}]{carleo2012localization}%
  \BibitemOpen
  \bibfield  {author} {\bibinfo {author} {\bibfnamefont {G.}~\bibnamefont
  {Carleo}}, \bibinfo {author} {\bibfnamefont {F.}~\bibnamefont {Becca}},
  \bibinfo {author} {\bibfnamefont {M.}~\bibnamefont {Schir{\'o}}},\ and\
  \bibinfo {author} {\bibfnamefont {M.}~\bibnamefont {Fabrizio}},\ }\bibfield
  {title} {\bibinfo {title} {Localization and glassy dynamics of many-body
  quantum systems},\ }\href@noop {} {\bibfield  {journal} {\bibinfo  {journal}
  {Scientific reports}\ }\textbf {\bibinfo {volume} {2}},\ \bibinfo {pages}
  {243} (\bibinfo {year} {2012})}\BibitemShut {NoStop}%
\bibitem [{\citenamefont {Biroli}\ and\ \citenamefont
  {Tarzia}(2017)}]{biroli2017delocalized}%
  \BibitemOpen
  \bibfield  {author} {\bibinfo {author} {\bibfnamefont {G.}~\bibnamefont
  {Biroli}}\ and\ \bibinfo {author} {\bibfnamefont {M.}~\bibnamefont
  {Tarzia}},\ }\bibfield  {title} {\bibinfo {title} {Delocalized glassy
  dynamics and many-body localization},\ }\href@noop {} {\bibfield  {journal}
  {\bibinfo  {journal} {Physical Review B}\ }\textbf {\bibinfo {volume} {96}},\
  \bibinfo {pages} {201114} (\bibinfo {year} {2017})}\BibitemShut {NoStop}%
\bibitem [{\citenamefont {Vojta}(2019)}]{vojta2019disorder}%
  \BibitemOpen
  \bibfield  {author} {\bibinfo {author} {\bibfnamefont {T.}~\bibnamefont
  {Vojta}},\ }\bibfield  {title} {\bibinfo {title} {Disorder in quantum
  many-body systems},\ }\href@noop {} {\bibfield  {journal} {\bibinfo
  {journal} {Annual Review of Condensed Matter Physics}\ }\textbf {\bibinfo
  {volume} {10}},\ \bibinfo {pages} {233} (\bibinfo {year} {2019})}\BibitemShut
  {NoStop}%
\bibitem [{\citenamefont {Ross}(2014)}]{ross2014first}%
  \BibitemOpen
  \bibfield  {author} {\bibinfo {author} {\bibfnamefont {S.~M.}\ \bibnamefont
  {Ross}},\ }\bibfield  {title} {\bibinfo {title} {A first course in
  probability},\ }\href@noop {} {\bibfield  {journal} {\bibinfo  {journal}
  {Pearson}\ } (\bibinfo {year} {2014})}\BibitemShut {NoStop}%
\bibitem [{\citenamefont {Chandrasekhar}(1943)}]{chandrasekhar1943stochastic}%
  \BibitemOpen
  \bibfield  {author} {\bibinfo {author} {\bibfnamefont {S.}~\bibnamefont
  {Chandrasekhar}},\ }\bibfield  {title} {\bibinfo {title} {Stochastic problems
  in physics and astronomy},\ }\href@noop {} {\bibfield  {journal} {\bibinfo
  {journal} {Reviews of Modern Physics}\ }\textbf {\bibinfo {volume} {15}},\
  \bibinfo {pages} {1} (\bibinfo {year} {1943})}\BibitemShut {NoStop}%
\bibitem [{\citenamefont {Kalikow}(1981)}]{kalikow1981poisson}%
  \BibitemOpen
  \bibfield  {author} {\bibinfo {author} {\bibfnamefont {S.}~\bibnamefont
  {Kalikow}},\ }\bibfield  {title} {\bibinfo {title} {A poisson random walk is
  bernoulli},\ }\href@noop {} {\bibfield  {journal} {\bibinfo  {journal}
  {Communications in Mathematical Physics}\ }\textbf {\bibinfo {volume} {81}},\
  \bibinfo {pages} {495} (\bibinfo {year} {1981})}\BibitemShut {NoStop}%
\bibitem [{\citenamefont {Das}\ \emph {et~al.}(2019)\citenamefont {Das},
  \citenamefont {Mal}, \citenamefont {Sen(De)},\ and\ \citenamefont
  {Sen}}]{das2019inhibition}%
  \BibitemOpen
  \bibfield  {author} {\bibinfo {author} {\bibfnamefont {S.}~\bibnamefont
  {Das}}, \bibinfo {author} {\bibfnamefont {S.}~\bibnamefont {Mal}}, \bibinfo
  {author} {\bibfnamefont {A.}~\bibnamefont {Sen(De)}},\ and\ \bibinfo {author}
  {\bibfnamefont {U.}~\bibnamefont {Sen}},\ }\bibfield  {title} {\bibinfo
  {title} {Inhibition of spreading in quantum random walks due to quenched
  poisson-distributed disorder},\ }\href@noop {} {\bibfield  {journal}
  {\bibinfo  {journal} {Physical Review A}\ }\textbf {\bibinfo {volume} {99}},\
  \bibinfo {pages} {042329} (\bibinfo {year} {2019})}\BibitemShut {NoStop}%
\bibitem [{\citenamefont {Watabe}\ \emph {et~al.}(2008)\citenamefont {Watabe},
  \citenamefont {Kobayashi}, \citenamefont {Katori},\ and\ \citenamefont
  {Konno}}]{watabe2008limit}%
  \BibitemOpen
  \bibfield  {author} {\bibinfo {author} {\bibfnamefont {K.}~\bibnamefont
  {Watabe}}, \bibinfo {author} {\bibfnamefont {N.}~\bibnamefont {Kobayashi}},
  \bibinfo {author} {\bibfnamefont {M.}~\bibnamefont {Katori}},\ and\ \bibinfo
  {author} {\bibfnamefont {N.}~\bibnamefont {Konno}},\ }\bibfield  {title}
  {\bibinfo {title} {Limit distributions of two-dimensional quantum walks},\
  }\href@noop {} {\bibfield  {journal} {\bibinfo  {journal} {Physical Review
  A}\ }\textbf {\bibinfo {volume} {77}},\ \bibinfo {pages} {062331} (\bibinfo
  {year} {2008})}\BibitemShut {NoStop}%
\end{thebibliography}


\providecommand{\noopsort}[1]{}\providecommand{\singleletter}[1]{#1}%
%


\end{document}